# Instabilities in the wake of an inclined prolate spheroid

**Helge I. Andersson, Fengjian Jiang, Valery L. Okulov**

**ABSTRACT**: We investigate the instabilities, bifurcations and transition in the wake behind a 45-degree inclined 6:1 prolate spheroid, through a series of direct numerical simulations (DNS) over a wide range of Reynolds numbers ($Re$) from 10 to 3000. We provide a detailed picture of how the originally symmetric and steady laminar wake at low $Re$ gradually looses its symmetry and turns unsteady as $Re$ is gradually increased. Several fascinating flow features have first been revealed and subsequently analysed, *e.g.* an asymmetric time-averaged flow field, a surprisingly strong side force *etc*. As the wake partially becomes turbulent, we investigate a dominating coherent wake structure, namely a helical vortex tube, inside of which a helical symmetry alteration scenario was recovered in the intermediate wake, together with self-similarity in the far wake.

## 1. Introduction

The interest in flows of real (*i.e.* viscous) fluids around spheroidal bodies can be traced back for more than a century. Oberbeck (1876) formulated an integral equation for flow around a prolate spheroid and obtained the drag force on the spheroid when the major (*i.e.* symmetry) axis is aligned with the flow direction. In the 1920s, Jeffery (1922) derived equations for the torques exerted on a spheroidal body in a steady but otherwise arbitrary flow field in the creeping-flow limit, *i.e.* zero Reynolds number. Jeffery's pioneering work was later followed by many others, notably Breach (1961) and Brenner (1964) in the 1960s, the latter who derived analytical expressions for the forces that act on an arbitrarily oriented spheroid in creeping flow.

**Helge I. Andersson**
Department of Energy and Process Engineering, Norwegian University of Science and Technology (NTNU), 7491 Trondheim, Norway
e-mail: helge.i.andersson@ntnu.no

**Fengjian Jiang** (✉)
Department of Marine Technology, NTNU, 7491 Trondheim, Norway
e-mail: fengjian.jiang@ntnu.no

**Valery L. Okulov**
Department of Wind Energy, Technical University of Denmark, 2800 Lyngby, Denmark
Institute of Thermophysics, SB of RAS, 630090 Novosibirsk, Russia
e-mail: vaok@dtu.dk

A major advantage of the works by Jeffery (1922) and Brenner (1964) is that their results are valid for *prolate* spheroids with aspect ratio $\lambda > 1$ as well as for *oblate* spheroids with aspect ratio $\lambda < 1$. The aspect ratio is defined herein as the ratio between the length $2c$ of the symmetry axis and the length $2a$ of the other two axes. Brenner's expression for the force on a spheroid also includes the Stokes drag force for $\lambda = 1$, *i.e.* for creeping flow around a sphere. Nowadays, Brenner-forces and Jeffery-torques are routinely used in simulations of particle-laden flows in a Lagrangian-Eulerian point-particle approach; see *e.g.* the recent review by Voth & Soldati (2017).

Most of the early studies focused on the forces and torques on the spheriodal geometry. The flow field around the spheroid remained essentially untouched until the 1970s. Since then, however, several studies have considered the flow around a spheroidal body, and eventually also the wake behind it, by means of both experimental and computational methods. Through the latest half-century, the spheroid has emerged as a perfect idealization for explorations of complex three-dimensional flow separations and bluff-body wake phenomena. However, compared to the long history of investigations of cylinder wakes (Zdravkovich 1997), the current knowlegde of the wake behind inclined spheroids is still far from comprehensive. The 50-year period in which flows around and behind spheroids have been actively researched and exploited can be roughly divided into four different stages, each in which the research was focussed differently, partly driven by advances of the available research tools, *e.g.* optical measurement techniques or the advent of high-performance computing (HPC).

**The 1$^{st}$ period** from the early 1970s to the late 1980s is characterized by observations of the surface flow on the spheroid, by means of numerical simulations or experimental measurements. Two different separation models, originally proposed by Maskell (1955), were first introduced by Wang (1972) to flow around a spheroid, and named *open separation* and *closed separation*. Three years later, Wang (1975) performed a more detailed investigation on flow separation, and his view was well accepted by the community. Since the open- and closed-separation idea was introduced and until the end of 1980s, extensive investigations have been carried out. In experimental studies, the observations made by Peake & Tobak (1982) and Wang *et al.* (1990) more convincingly showed the existence of the two different separation types. Costis *et al.* (1989) showed that the two types can co-exist in flow around a spheroid at certain attack angles.

In this period, experimental studies mainly focused on observations of the surface flow on the spheroid. Noteworthy examples are the wall shear-stress distribution observed by Meier & Kreplin (1980) and the limiting streamlines reported by Han & Patel (1979) and Costis *et al.* (1989). Meanwhile, several experiments made efforts to measure the velocity and vorticity field very close to the spheroid, see for example Costis & Telionis (1984). These measurements were usually made by laser sheets, particle displacement velocimetry (PDV) or Laser-Doppler velocimetry (LDV).

Compared to the experimental studies, the outcome of computer simulations is more fruitful. However, inasmuch as the computers were limited in terms of speed and memory, nearly all the simulations solved the boundary layer equations instead of the full Navier-Stokes (N-S) equations in order to obtain the flow field very close to the spheroid. Blottner and Ellis (1973), Geissler (1974), and Wang (1974) carried out very early numerical studies in this field by solving the boundary layer equations. Later, Cebeci *et al.* (1981), Ragab (1982), and Patel & Baek (1985) followed similar computational approaches and solved the boundary layer flow around spheroids with different aspect ratios and inclination angles.

Although all details of the flow field could not be obtained by solving the boundary layer equations, these studies nevertheless contributed to the understanding of the two different separation models, as well as the effects caused by the aspect ratio and attack angle. The numerical results moreover served to validate experimental observations, and even had the ability to provide more information than what was achieved in the experiments, *e.g.* the velocity distribution in the viscous boundary layer.

It is noteworthy that in this period, be it in experimental or numerical studies, 4:1 and 6:1 prolate spheroids were the most popular geometries; whereas allmost all studies focussed on inclination angles lower than $30^\circ$ with respect to the inflow direction.

**The 2$^{nd}$ period** extended from the late 1980s to around 2000. Experimentalists became able to measure the near-wake behind different spheroids, while numerical investigators left the boundary layer equations and instead solved simplified N-S equations.

As the capacity of modern computers rapidly developed, the limitations of the boundary layer equations were avoided by instead solving simplified N-S equations. Pan & Pulliam (1986), Cebeci & Su (1988), Vasta *et al.* (1989), and Patel & Kim (1994) carried out numerical simulations of the flow around a spheroidal body by simplifying the full N-S equations in some different ways. Although these studies focused more on the numerical methods than on the flow physics, the computed solutions also provided new insight in the three-dimensional flow around an inclined spheroid. The probably first numerical solutions of the full N-S equations were reported in 1987 by Shirayama & Kuwahara (1987). They considered the flow around a 2:1 prolate spheroid for Re = 100 and at several different angles of attack up to $45°$ and even for Re = 1000 at attack angle $15°$. Although the accuracy of their computations might be questioned due to the fairly coarse grid, their flow visualization plots are indeed fascinating.

Different from in the 1$^{st}$ period, experimental studies somewhat dominated the research activities in this 2$^{nd}$ period. A team at Virginia Tech., headed by Professor R.

L. Simpson, adopted in the early 1990s the prolate spheroid as an adequate body on which to study three-dimensional flow separation. Alongside with researchers elsewhere, they contributed many important observations and measurements, some of which are still popular validation cases for computational fluid dynamics (CFD) codes.

Kreplin & Stager (1993) measured the Reynolds stress in the separated boundary layer around a spheroid, and offered data useful for validation of turbulence models. Chesnakas & Simpson (1994) used an improved LDV method to measure the separation point on a 6:1 prolate spheroid under $10°$ inclination angle and compared with earlier measurements by Barber & Simpson (1991). Ahn and Simpson (1992), Wetzel & Simpson (1993), Hoang *et al.* (1994a, b), and Fu *et al.* (1994) measured the unsteady flow around spheroids with different aspect ratios and attack angles, mostly focussing on the three-dimensional flow separation. In the late 1990s, Chesnakas & Simpson (1997) further improved their LDV measurement technique and managed to obtain the time-averaged flow field around a 6:1 prolate spheroid. Wetzel *et al.* (1998) first observed features of cross-flow separation in an inclined spheroid flow, and thereafter accurately measured the separation point on a spheroid under unsteady motion (Wetzel & Simpson, 1998). Goody *et al.* (1998) measured the velocity spectra in the wake of a spheroid having a small inclination angle. Goody *et al.* (2000) and Goody & Simpson (2000) investigated the surface pressure fluctuations on a prolate spheroid.

Extensive experimental measurements were performed and fruitful novel results were obtained in this period. The availability of these experimental data made the prolate spheroid, especially that with aspect ratio 6:1, become one of the most popular validation cases for different turbulence closure models requiered in CFD-codes based on the Reynolds-averaged Navier-Stokes (RANS) equations. Most of the experiments in this period utilized LDV-methods at Reynolds numbers ~ $10^6$. Further information on experimental investigatiosn of flow around spheroids is provided in the two review papers by Simpson (1996, 2005).

**The 3rd period,** the decade from around 2000 to around 2010, during which experimental activities were reduced and the prolate spheroid became a prominent validation case for various turbulence models. A rapid development of computer capacity and speed was witnessed as the 21st century emerged and CFD gained popularity in both fundamental and engineering research. Improved turbulence models made it possible for researchers and engineers to investigate a variety of fairly complex flow problems, among which the flow around a prolate spheroid became an attractive validation case, partly due to its complex flow features (in spite of its geometrical simplicity), and partly due to the abundance of experimental results accumulated during the 2nd period.

Rhee & Hino (2000; 2002) used a one-equation turbulence model to simulate the flow around a prolate spheroid. They compared the computed results with previous experimental data and concluded that the one-equation model was only able to qualitatively capture some important flow features. Cummings *et al.* (2003) continued to investigate this issue and suggested explanations for the quantitative inaccuracies. Taylor *et al.* (2005) used a prolate-spheroid-like submarine model to study the performance and realism of some different turbulence closures, and concluded that some models gave better force predictions than others. It was soon realized that although RANS-based approaches provide reasonably good integrated forces on different bluff-body geometries, they cannot produce reliable details of the unsteady three-dimensional flow field.

Along with the further advancement of the computer technology, more advanced approaches to handle turbulent flows came into use. Alin *et al.* (2005) pointed out that, to more accurately simulate the velocity field around a spheroid, one should use URANS (Unsteady RANS) or LES (Large-Eddy Simulation) rather than steady RANS-based approaches. Constantinescu *et al.* (2002) and Kotapati-Apparao *et al.* (2003) adopted the flow around a spheroid to validate DES (Detached Eddy Simulations), whereas Wikström et al. (2004) and Karlsson & Fureby (2009) used LES to investigate this flow problem.

Numerous other computational studies were also conducted in this period, the majority of which focused either on validation of different turbulence models or on numerical codes, rather than on the fascinating flow physics. In spite of the many publications that emerged in this period, the understanding of the complex flow physics made only modest advancements. Until the end of this 3$^{rd}$ period, the vast majority of research efforts concentrated on the flow field in the close vicinity of the spheroidal geometry and no in-depth studies of the wake flow were reported.

**The 4$^{th}$ period** from around 2010 and until this day, witnessed explorations of fundamental flow physics by means of direct numerical simulations (DNS) of wakes behind prolate spheroids under different attack angles.

DNSs provide detailed flow field information, more than any other CFD-tools, and comparable to or even more than that in experiments. Together with advanced flow visualization techniques, numerically generated wake flow data can be presented in almost whatever way one may want to. Unfortunately, however, to perform DNS of wakes behind spheroids is more demanding than DNS of most canonical flow problems, *e.g.* turbulent channel and pipe flows, turbulent boundary layers, and homogeneous isotropic turbulence. This is because a wake simulation requires a relatively large computational domain in combination with a refined computational grid, not only in the vicinity of the surface of the spheroid but also in the downstream wake. Thanks to the continuous and rapid development of high-performance computers, the use of the DNS approach to wake flow behind prolate spheroids has now become feasible.

The first detailed DNS investigation of the flow around a spheroid was presented by El Khoury *et al.* (2010), who simulated the wake behind a 6:1 prolate spheroid at Re = 10 000 under 90º inclination angle, *i.e.* in cross-flow. The computated results showed interesting wake dynamics. A systematic DNS-study of the wake development at substantially lower Reynolds numbers was reported by El Khoury *et al.* (2012). The main features of the wake behind the 6:1 spheroid in cross-flow turned out to be distinctly different from both the cylinder wake and the sphere wake. Jiang *et al.* (2014; 2015b) carried out a DNS investigation of the wake behind the same 6:1 prolate spheroid but inclined 45º with respect to the oncoming flow. They showed, among other things, that the wake behind the 45º inclined spheroid is fundamentally different from that in the cross-flow orientation. These DNS studies demonstrated the striking effects brought about by the attack angle even though the aspect ratio remained the same.

The scope of the present study is to summarize the current understanding of the instabilities and bifurcations in the wake behind one particular configuration, namely the 45º inclined 6:1 prolate spheroid. The main features of the wake will be carefully described, including how the originally laminar wake first becomes unsteady and eventually transits to turbulence. At higher Reynolds numbers, when the wake is already turbulent, we will focus mostly on the coherent vortical wake structures. The results from the different flow regimes are presented in Sections 3, 4, and 5.

## 2. Governing Equations and Computational Methods

### 2.1 Flow problem

In this study, we focus on the wakes behind a 45-degree inclined prolate spheroid, whose aspect ratio $\lambda = c/a$ is 6:1. We consider a range of Reynolds numbers from 10 to 3000, which covers a wide range of flow regimes. The Reynolds number is based on the free-stream velocity $U_0$ and the length $D = 2a$ of the minor axis of the spheroid, *i.e.* $Re_D = U_0 D/\nu$.

The stream-wise direction is referred to as the *x*-direction, the vertical direction as the *y*-direction, and the cross-flow direction as the *z*-direction. Note that 'vertical' is meant as upward in the plots, and does not imply that a gravity force is considerd. In addition, to conveniently present and discuss results in the near-wake around the inclined spheroid, we introduced a fixed-to-body coordinate system ($\xi, \eta, z$). The origins of both coordinate systems coincide in the geometrical center of the spheroid, *i.e.* at the intersection between the major and minor axes. The flow configuration in the symmetry plane, *i.e.* the (*x*, *y*)-plane at $z/D = 0$, is depicted in Fig. 1(a), where both coordinate systems are shown in the inset panel (b).

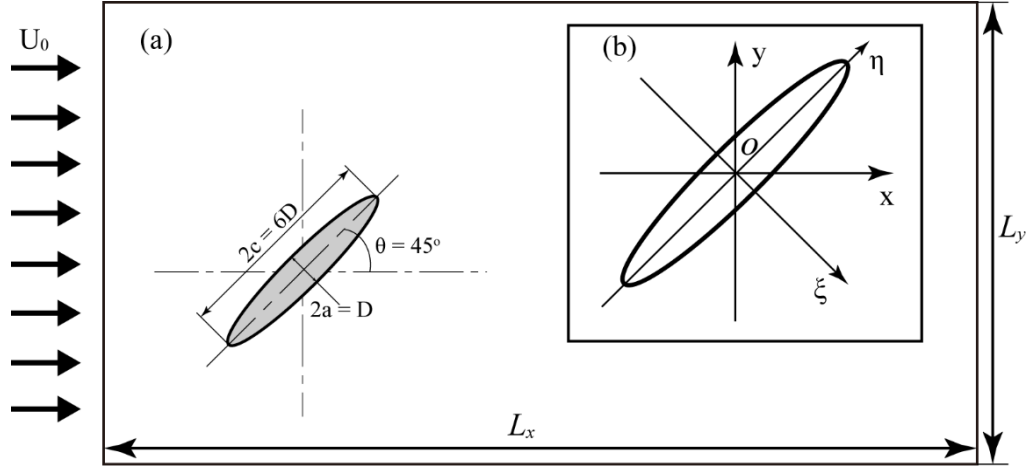

**Fig. 1** (a) The flow configuration in the symmetry plane, *i.e.* the ($x$, $y$)-plane at $z = 0$. (b) The two coordinate systems ($x$, $y$, $z$) and ($\xi$, $\eta$, $z$). The computational domain is a three-dimensional rectangular box, with lengths $L_x$, $L_y$ and $L_z$ in the corresponding coordinate directions.

## 2.2 Governing equations and numerical methods

All DNSs in this study were conducted with a second-order finite-volume Navier-Stokes solver MGLET (Manhart *et al.* 2001; Manhart 2004). The full three-dimensional Navier-Stokes equations (in integral form) for incompressible fluids:

$$\oint_A \boldsymbol{u} \cdot \boldsymbol{n}\, dA = 0 \tag{1}$$

$$\frac{\partial}{\partial t}\iiint_\Omega u_i\, d\Omega + \oint_A u_i \boldsymbol{u} \cdot \boldsymbol{n}\, dA = -\frac{1}{\rho}\oint_A p\boldsymbol{i_i} \cdot \boldsymbol{n}\, dA + \nu \oint_A grad\, u_i \cdot \boldsymbol{n}\, dA \tag{2}$$

are discretized on a three-dimensional Cartesian grid. In equation (1) and (2), $\Omega$ denotes the control volume, $A$ represents the control surface, $\boldsymbol{n}$ is the unit vector on $dA$ pointing out of the surface, $\boldsymbol{i_i}$ is the unit vector in the $x_i$-direction, while $\boldsymbol{u}$ is the velocity vector in a Cartesian coordinate system. MGLET uses a staggered grid arrangement, which means that the pressure nodes (centered at the center of the control volume) and the velocity nodes (centered at the center of control volume faces) are not co-located.

Each of the terms in the momentum equation (2) are discretized onto the staggered Cartesian grid. The surface integrals are approximated by the mid-point rule, while the volume integral is approximated by the volume of the control volume times the value of the integrand. The derivative in the diffusive term is approximated with a central-difference formulation, such that the overall second-order accuracy is fulfilled. The time integration of the discretized governing equations is achieved by an explicit low-storage 3rd-order Runge-Kutta scheme (Williamson 1980), during which the

Poisson equation is solved with Stone's (Stone 1968) strongly implicit procedure (SIP), in order to correct the pressure and assure a divergence-free velocity field.

A particular challenge is the representation of the solid surface of the spheroid in the staggered Cartesian grid arrangement. This is handled by a direct-forcing immersed boundary method (IBM). In MGLET, the surface of the object, *i.e.* the spheroid in this paper, is mapped by triangular cells, under the widely used STL (stereolighograph) standards. The resolution of the surface mapping is controlled by the user. In the present study, the surface mapping reached a tolerance smaller than $0.001D$, *i.e.* substantially smaller than the smallest grid size $0.006D$ used in the $Re_D = 3000$ simulations.

The IBM method in MGLET starts to find the intersection points of the grid lines and the body surface, which means no Lagrangian markers need to be pre-defined for the body surface. The searching strategy is pressure cell oriented. When an intersection point is found, and lie within the pressure cell, this cell will be blocked. Then each velocity cell staggered with the blocked pressure cell is also blocked. The "blocking" process is essentially a process to enforce Dirichlet boundary conditions to the intersected cells by interpolating flow variables from the surrounding neighbour cells. It is worth mentioning that MGLET employs a flux correction at the fluid-solid interface, which assures mass conservation. Further details about the IBM-method in MGLET, including a stability analysis, can be found in Peller et al. (2006) and Peller (2010).

**Table 1**. Computational information for cases at different $Re_D$

| $Re_D$ | Domain size $L_x \times L_y \times L_z$ | Minimum Grid spacing | Grid cells $N_x \times N_y \times N_z$ | Total Grid numbers |
|---|---|---|---|---|
| 50 | $28D \times 32D \times 11D$ | $0.04D$ | $320 \times 340 \times 120$ | $13.06 \times 10^6$ |
| 200 | $28D \times 32D \times 11D$ | $0.03D$ | $400 \times 420 \times 184$ | $30.24 \times 10^6$ |
| 800 | $38D \times 24D \times 11D$ | $0.01D$ | $900 \times 708 \times 292$ | $18.17 \times 10^7$ |
| 1000 | $38D \times 24D \times 11D$ | $0.01D$ | $900 \times 708 \times 292$ | $18.17 \times 10^7$ |
| 1200 | $38D \times 24D \times 11D$ | $0.01D$ | $900 \times 708 \times 292$ | $18.17 \times 10^7$ |
| 3000 | $38D \times 24D \times 21D$ | $0.006D$ | $1344 \times 1024 \times 544$ | $74.87 \times 10^7$ |

**2.3 Grid independence study**

In our earlier paper by Jiang et al. (2015b), a detailed grid-independence study was presented for $Re_D = 3000$. Three different minimum grid sizes Δ were chosen to generate three meshes, *i.e.* the coarse mesh with Δ =$0.012D$, the medium mesh with Δ = $0.008D$, and the fine mesh with Δ = $0.006D$, respectively. The computed drag forces obtained for the three meshes were compared, and the deviation between the medium and fine mesh was only 0.4%. Moreover, a comparison of the time-averaged flow fields clearly indicated a grid-converging trend. The finest mesh with $\Delta = 0.006D$ is therefore believed to be sufficiently fine for all the DNS studies. At lower Reynolds numbers, similar grid independence studies were also conducted, as thoroughly discussed in Jiang et al. (2014, 2015a), the details of which will not be repeated herein.

In table 1, the most essential computational parameters used for the different $Re$-cases is summarized. Notice in particular that the size of the computational domain varies with Reynolds number. It should also be mentioned that only data corresponding to the finest mesh at each $Re_D$ are included in table 1.

## 3. Low Reynolds Number Laminar Wake

In this Section we consider the low-Reynolds-number range in which the flow around the inclined prolate spheroid is steady and symmetric about the geometrical mid-plane, *i.e.* about $z = 0$. A particularly useful treatise on the general subject of low-Re flows has been provided by Happel & Brenner (1973).

### 3.1 Very low Reynolds number ($Re_D \leq 10$)

In the extreme limit as $Re \rightarrow 0$, the non-linear inertial terms in the Navier-Stokes equation vanish and we are only left with a balance between the terms associated with pressure and the viscous stresses. While the influence of the viscous stresses is confined to thin boundary layers at high Reynolds numbers ($Re \gg 1$), the viscous stresses are influential even far away from the body for $Re \sim 1$ and affect the flow field all the way to infinity in the creeping-flow limit ($Re = 0$). The linearity of the limiting momentum equation made it possible to derive analytical expressions for the force (Brenner, 1964) and torque (Jeffery, 1922) exerted by the viscous fluid on a spheroidal body at an arbitrary angle of attack.

The force and torque expressions were adopted by Zhang *et al.* (2001) and Mortensen *et al.* (2008) to simulate the behaviour of numerous prolate spheroids suspended in a turbulent flow. In this so-called point-particle approach, the individual spheroids translate and rotate in accordance with the viscous forces and torques. Although the point-particle approach was first developed for tiny spherical particles, see *e.g.* Eaton (2009), the recent overviews by Andersson & Soldati (2013) and Voth & Soldati (2017) illustrate how the approach has been further developed and adopted also for non-spherical particles, notably spheroids.

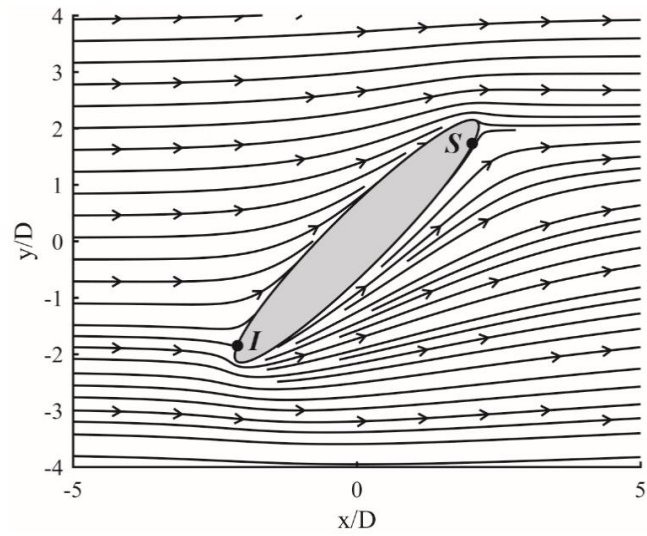

Fig. 2 Streamlines in the symmetry plane ($z/D = 0$); $Re_D = 10$.

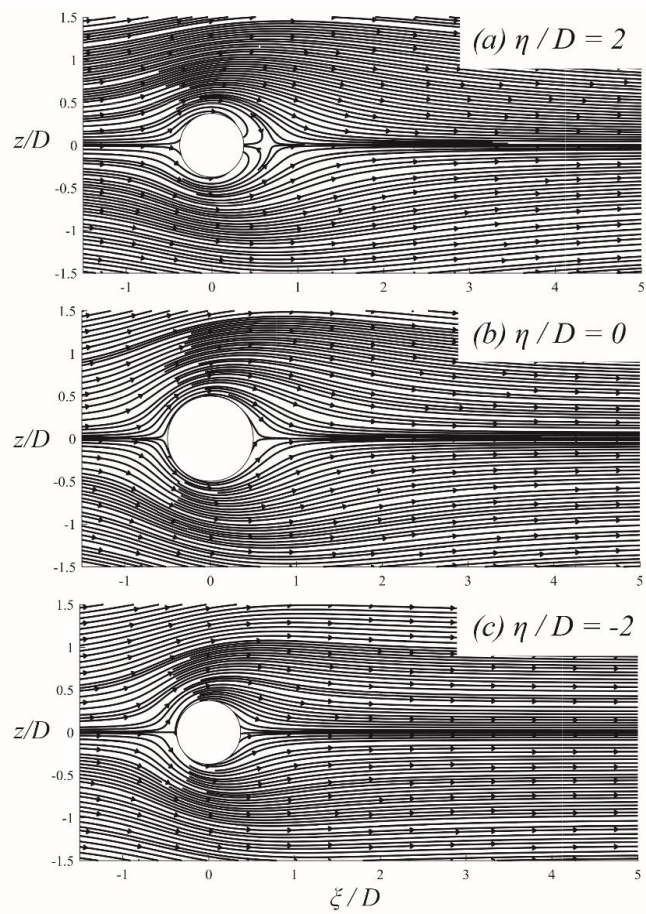

Fig. 3 Streamlines of projected velocity vector on three different cross-sectional ($\xi$, $z$)-planes; $Re_D = 10$.

A limitation of the point-particle approach is that the expressions for the force and torque are formally valid only for *Re* = 0. However, the Stokes drag on a sphere is a good approximation at *Re* = 0.1 and nearly up to *Re* = 1. The same reasoning is carried over also to spheroidal particles, although without any rigorous proof of evidence. Several semi-empirical formulas for the drag force on a sphere have been published over the years, of which the simple expression by Schiller & Naumann (1933) is among the most popular. The Schiller-Naumann correlation is known to provide an accurate drag force up to *Re* ≈ 800. Motivated by the need for extensions of the force and torque expressions for spheroids to finite Reynolds numbers, some alternative formulas have been developed during the last decade. These formulas were primarily obtained by curve-fitting to computed forces and torques, notably by Hölzer & Sommerfeld (2009), Zastawny *et al.* (2012) and Ouchene *et al.* (2016). Sanjeevi & Padding (2017) explored the orientational dependence of drag on spheroids over a fairly large range of Reynolds numbers. In our earlier paper (Jiang *et al.* 2014), we showed computed results for drag, lift and torques for a 6:1 prolate spheroid at 45° attack at *Re* = 50 and compared with the Hölzer & Sommerfeld (2008) and Zastawny *et al.* (2012) correlations.

A challenge of CFD at low Reynolds numbers is that the viscous effects spread far away from the bluff body. The common usage of free-slip boundary conditions, *i.e.* Neumann conditions of the tangential velocity components and Dirichlet condition on the normal velocity component, on the sides of the computational domain cannot be justified unless the cross-section of the computational domain is unusually large. The need for an exceptionally large cross-section, which inevitably is computationally expensive, increases as the Reynolds number decreases.

Here, as an example, we show results for $Re_D$ = 10. The streamlines in the geometrical symmetry plane (*z/D* = 0) in Fig. 2 show that the flow impinges near the leading pole of the spheroid. The flow is directed along both sides of the spheroid until the flow along the rear side separates from the curved surface close to the upper pole. The separation point (*S*) is located almost antisymmetrically of the impingement point (*I*) at this fairly low but yet finite *Re*. A perfect fore-aft symmetry is expected only at *Re* = 0. Streamlines of projected velocity vectors in cross-sectional (*ξ, z*)-planes are shown in Fig. 3. One should keep in mind that there is a velocity component perpendicular to the projection plane, therefore the "streamlines" in Fig. 3 are not true streamlines. Yet these "streamlines of the projected velocity vector" represent a practical way to visualize the flow field, which offers useful information to interpret and analyze the results. A similar visualization approach has also been used in earlier publications, see *e.g.* Johnson & Patel (1999).

In the equatorial plane (*η* = 0), a modest fore-aft asymmetry can be observed, similarly as for the two-dimensional flow field around an infinitely long circular cylinder at a *Re* only slightly above 1, see *e.g.* the streamline pattern at *Re* = 1.54 in Van Dyke (1982). The flow field apparently depends on the location along the symmetry axis of the spheroid. The streamlines 2D below the equatorial plane suggests that the flow is on the verge to separate from the surface and resembles the flow field around a circular cylinder at a Reynolds number slighly below unity. The flow topology 2*D* above the equatorial plane is rather different and exibits a distinct saddle point in the wake. The distinctly different flow patterns at $\eta = -2D$ and $\eta = +2D$ is an effect of the finite Reynolds number.

### 3.2  $Re_D$ = 50 and 200

The computed flow fields at $Re_D = 50$ and 200 remain steady, laminar and symmetric about the (*x, y*)-plane at $z = 0$, just as for $Re_D = 10$. However, while the low-Re flow remained almost attached to the spheroid, a substantial wake emerges at the the two higher Reynolds numbers, as visualized in Fig. 4 by means of an iso-surface of $\lambda_2$, as suggested by Jeong and Hussain (1995). We observed that the laminar boundary layers around the spheroid at $Re_D = 50$ is thicker than those at $Re_D = 200$. The wake at $Re_D = 200$ is moreover significantly longer than that at $Re_D = 50$. Despite these differences, the two wakes in Fig. 4 share the same main features, namely that the separated vortex sheets from the two halves of the spheroid roll up to form a pair of counter-rotating vortices in the wake.

The separation of the boundary layers from the surface of the spheroid and the subsequent roll-up into a vortex pair is a consequence of the inclination of the spheroid relative to the oncoming flow which makes also the separation lines inclined. The vortex dynamics are governed by the vorticiy equation:

$$\frac{\partial \vec{\omega}}{\partial t} + (\vec{u} \cdot \nabla)\vec{\omega} = (\vec{\omega} \cdot \nabla)\vec{u} + \upsilon \nabla^2 \vec{\omega} \tag{3}$$

in which both non-linear advection $(\vec{u} \cdot \nabla)\vec{\omega}$ and vortex tilting $(\vec{\omega} \cdot \nabla)\vec{u}$ are essential mechanisms and contribute to the formation of the two counter-rotating vortex structures and their subsequent bending as they leave the shadow of the spheroid and become exposed to the strong streamwise velocity. The counter-rotating vortex pair remains almost aligned with the *x*-direction until the vorticity gradually fades away due to viscous diffusion $\upsilon \nabla^2 \vec{\omega}$. The symmetry of the wake at $Re_D = 50$ was retained also at $Re_D = 200$.

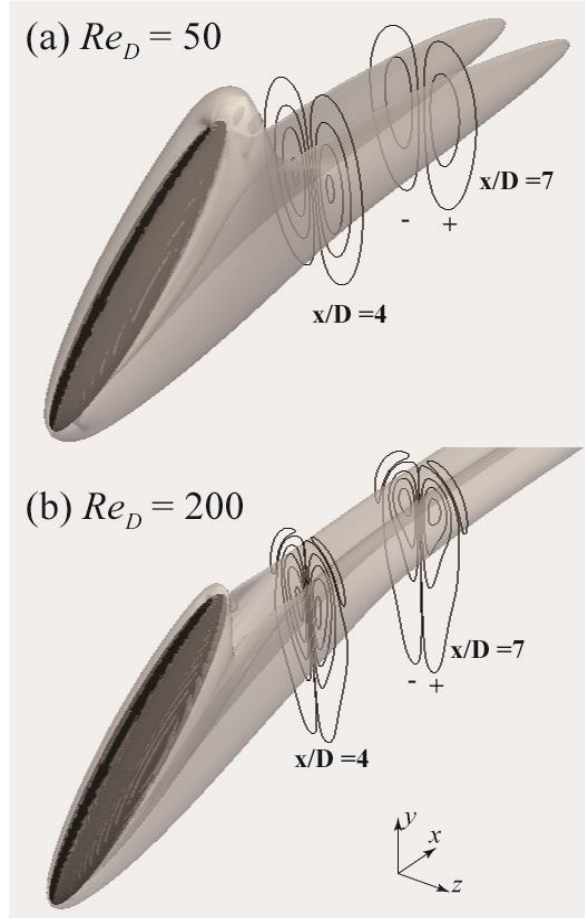

**Fig. 4** Perspective views of the wake structures at two different Reynolds numbers $Re_D$, visualized by means of an iso-surface of $\lambda_2$. (a) $Re_D = 50$, $\lambda_2 = -0.01$, together with streamwise vorticity ($\omega_x D/U_0$) contour lines in two x-slices, the values for the contour lines are in range (-0.4, 0.4). (b) Same as (a), but for $Re_D = 200$, $\lambda_2 = -0.1$, contour lines in range (-1.6, 1.6). The $\omega_x D/U_0$ distribution is anti-symmetric about the geometrical symmetry plane at $z = 0$ (the opposite signs are indicated).

## 4. Wake Transition and Initial Instability

### 4.1 The initial instability at $Re_D$ = 1000

The wake turns out to be different and more complex at $Re_D = 1000$, as can be seen in Fig. 5. The iso-surfaces of $\lambda_2 = -0.5$ in Fig. 5(a) and 5(b) show that the intermediate and far wake is no longer symmetric about the (x, y)-plane. In fact, while the left ($z < 0$) vortex tube maintains its coherence far downstream into the intermediate and far wake, the right ($z > 0$) vortex tube becomes distorted and partially disrupted some 10$D$ downstream. This observation is indeed confirmed by the plots of the vorticity magnitude in Fig. 5(c, d, e). While the vorticity is symmetric in the near-wake at x/D

= 4, a modest asymmetry becomes discernible at *x/D* = 8. Even further downstream, at *x/D* = 20, the two vortex tubes exhibit distinct asymmetries and only the left vortex has retained its tubular structure.

The force and torque coefficients defined as:

$$C_{Fi} = \frac{F_i}{\frac{1}{2}\rho U_0^2 \frac{\pi}{4} d^2} \; ; \; C_{Mi} = \frac{M_i}{\frac{1}{2}\rho U_0^2 \frac{\pi}{4} d^3} \; ; \; i = x, y, z \tag{4}$$

are computed and compiled in table 2 for several different $Re_D$. In the definitions (4), $F_i$ and $M_i$ are the $x_i$-components of the force vector $\vec{F} = (F_x, F_y, F_z)$ and the torque vector $\vec{M} = (M_x, M_y, M_z)$ which the viscous flow exerts on the spheroidal body. Here, *d* is the diameter of a volume-equivalent sphere, which for a 6:1 prolate spheroid becomes *d* = 1.817*D*. The data in table 2 clearly show that the side-force coefficient $C_{Fz}$ is negligibly small at $Re_D$ = 1000, just as at $Re_D$ = 50 and 200 when the wake is perfectly symmetric. This implies that, in spite of the asymmetries in the intermediate wake observed in Fig. 5, the near-wake characterized by the vortex sheets is still symmetric even at $Re_D$ = 1000. This conclusion is strongly supported by the observation that $C_{Mz}$ is the only non-negligible torque coefficient for all Reynolds numbers except $Re_D$ = 3000.

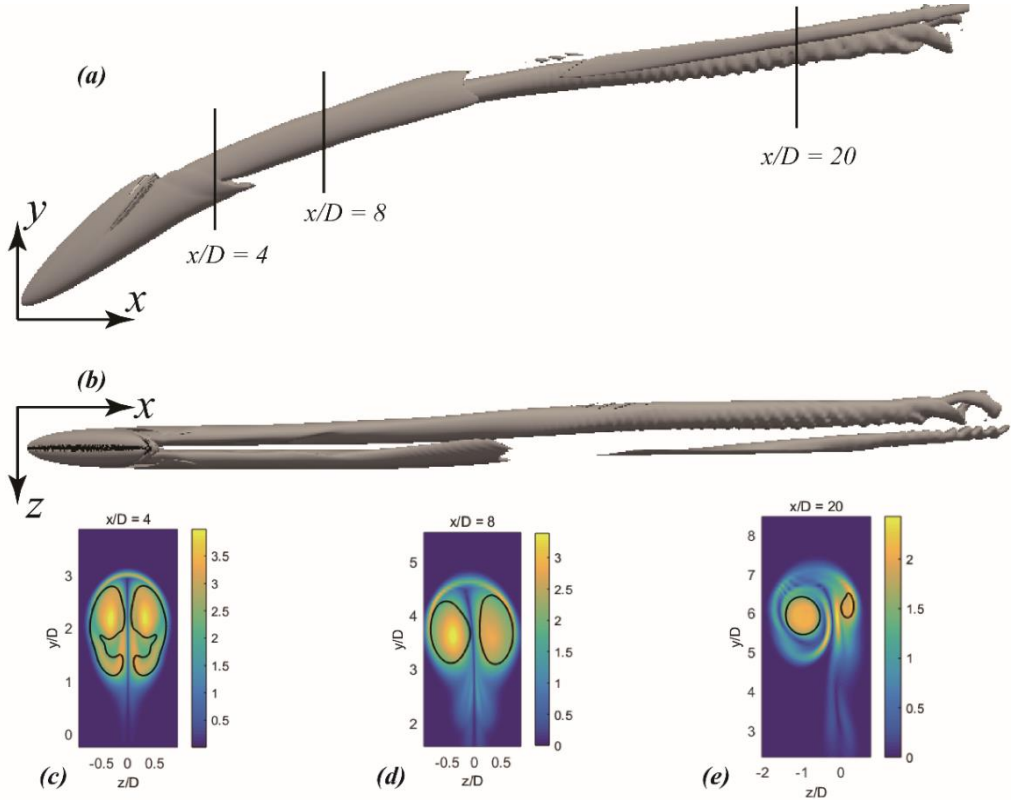

**Fig. 5** Overview of the wake at $Re_D$ = 1000. Iso-surface of $\lambda_2$ = -0.5; side view (a) and top view (b). The vorticity magnitude $|\omega|D/U_0$ is shown in vertical (*y, z*)-planes at *x/D* = 4, 8 and 20 in panels (c), (d) and (e), respectively. The black contour line in these three plots corresponds to $\lambda_2$ = -0.5.

**Table 2.** Force and torque coefficients at different Reynolds numbers $Re_D$

| $Re_D$ | $C_{Fx}$ | $C_{Fy}$ | $C_{Fz}$ | $C_{Mx}$ | $C_{My}$ | $C_{Mz}$ |
|---|---|---|---|---|---|---|
| 50 | 1.6915 | -0.7043 | -6.94E-04 | 7.17E-05 | 2.33E-04 | 0.5369 |
| 200 | 1.0406 | -0.6458 | -2.85E-04 | 2.46E-05 | 7.22E-05 | 0.4695 |
| 800 | 0.8254 | -0.6892 | -5.23E-05 | 1.84E-05 | -3.64E-05 | 0.4528 |
| 1000 | 0.8054 | -0.6984 | -9.25E-05 | 3.82E-05 | 5.09E-05 | 0.4483 |
| 1200 | 0.7897 | -0.7054 | 5.30E-05 | 9.54E-04 | -9.65E-04 | 0.4442 |
| 3000 | 0.8785 | -0.7959 | -0.6448 | -0.1684 | 0.1663 | 0.3109 |

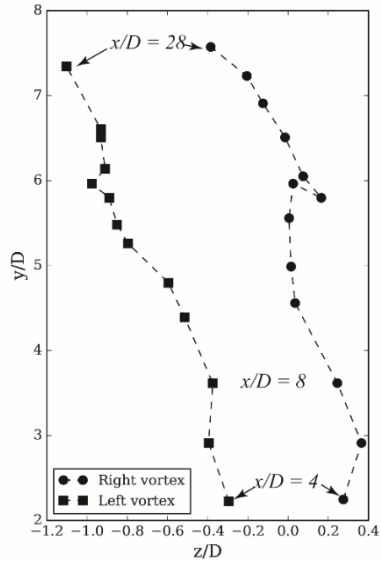

**Fig. 6** Traces of the center of each of the two vortex tubes. The center positions are determined by the local minimum of $\lambda_2$ in the vortex cross-sections traced at 13 different $x$-locations $2D$ apart from $x/D = 4$ to 28. $Re_D = 1000$.

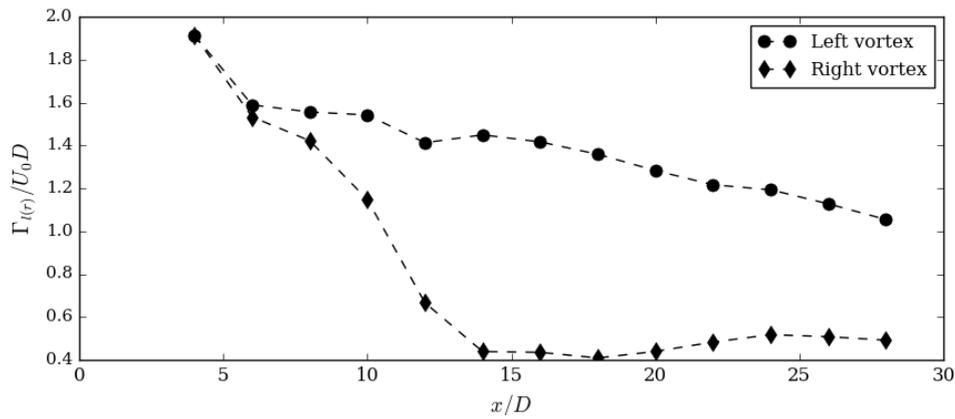

**Fig. 7** Circulation of the left ($\Gamma_l/U_0 D$) and right ($\Gamma_r/U_0 D$) vortex. $Re_D = 1000$.

Fig. 6 depicts the trajectories of the center of the two vortex tubes. The vortex tube center was defined by the local minimum of $\lambda_2$ in each vortex cross-section. From Fig. 6 we firstly observe that both vortex tubes are deflected to the $z < 0$ side.

This deflection is cleary seen also in the cross-sectional plot in Fig. 5(e). Combined with the information gained from Fig. 5, we also learn that the left-side vortex is more concentrated and coherent, whereas the right-side vortex deforms and starts to rotate around the left-side vortex after the symmetry between them is broken. From Fig. 5(b), one may even get the impression that the right-side vortex breaks up at around $x/D = 14$. This is, however, a visual artifact which arises because $\lambda_2 > -0.5$ in this region.

A rolled-up vortex is generally a rather stable vortical structure since the axial flow along the vortex axis induces vortex stretching, and the strong axial vorticity contributes to a compact core (Pradeep and Hussain, 2000). This can be seen also in the present case in Fig. 7, where the circulation of a vortex is calculated by integrating the streamwise vorticity over the vortex core (defined here as $\lambda_2 < -0.1$) in a given $x$-plane:

$$\Gamma_l = \left|\iint_{\Omega_l} \omega_x dydz\right|; \quad \Gamma_r = \left|\iint_{\Omega_r} \omega_x dydz\right| \tag{5}$$

Here, $\Omega_l$ and $\Omega_r$ refer to the vortex core region within the contour line $\lambda_2 = -0.1$. It should be mentioned that since the actual vortex axes are curves in three-dimensional space, the circulations calcaluated based on equation (5) are approximations of the circulation obtained by integration over a plane perpendicular to the vortex axis. However, since the misalignment between the vortex axes and the $x$-direction is modest, the expression in equation (5) is a fair approximation to the exact circulation.

From Fig. 7, we first observe that $\Gamma_l$ and $\Gamma_r$ are indistinguishable at $x/D = 4$, from which they start to diverge. The concentrated left-side vortex maintains high circulation almost all through the computational domain, which reveals a rather stable vortex. The circulation of the right-side vortex, however, decreases as soon as the symmetry is broken and reaches a minimum at about $x/D = 14$. This coincides with the observed break-up of the $\lambda_2$ iso-surface in Fig. 5(b).

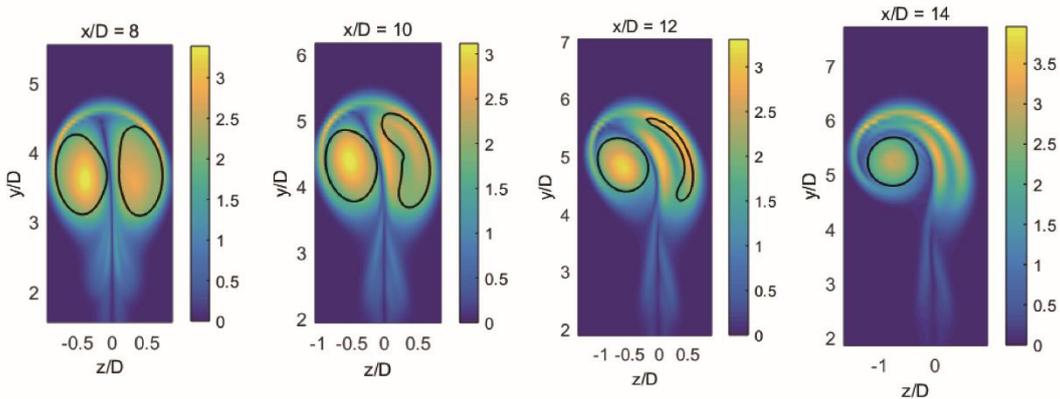

**Fig. 8** Distribution of vorticity magnitude $|\omega|D/U_0$ over four different slices at $x/D = 8$, 10, 12 and 14. The contour lines of $\lambda_2 = -0.5$ indicate the vortex cores. $Re_D = 1000$.

In Fig. 8, the distribution of the normalized vorticity magnitude $|\omega|D/U_0$ over the $(y, z)$-plane is shown at four different streamwise locations $x/D = 8$, 10, 12 and 14.

The counter-rotating vortex pair is fairly symmetric at $x/D = 8$, beyond which the left-side vortex gradually becomes more axisymmetric, visualized by the almost circular $\lambda_2$-contour, whereas the right-side vortex is bent into a boomerang and eventually totally deformed. Fig. 8 shows a manifestation of an elliptic instability of the right-side vortex core. However, the present scenario is different from the "cooperative elliptic instability" (Leweke and Williamson, 1998) of two long and concentrated counter-rotating vortices characterized by the coupled interaction between the two vortices rather than one dominating the other. This difference might originate from the close locations of the two vortex tubes in the present case, whereas the distance between two vortex centers was generally larger than 5 times the vortex size in the study by Leweke and Williamson (1998). Since the two vortices in Fig. 5 are very close, they influence each other directly through the strain field generated by themselves. As soon as the delicate balance between the two vortices is lost, the strain field generated by the stronger vortex (the left-side one in the present case) directly makes the weaker vortex rotate around it.

It should be noted that all the plots in Fig. 8 are time-independent, which means that they are representative both for the instantaneous as well as the time-averaged flow field. In fact, we can hardly observe any unsteadiness in this already asymmetric wake and all snapshots are almost identical at different instants of time. This suggests that the instability, although clearly visible, is at a very early stage and has not yet evolved into unsteadiness. One can easily imagine that if an unsteady strain field is generated by a concentrated vortex tube, the tube should rotate around itself with a certain frequency.

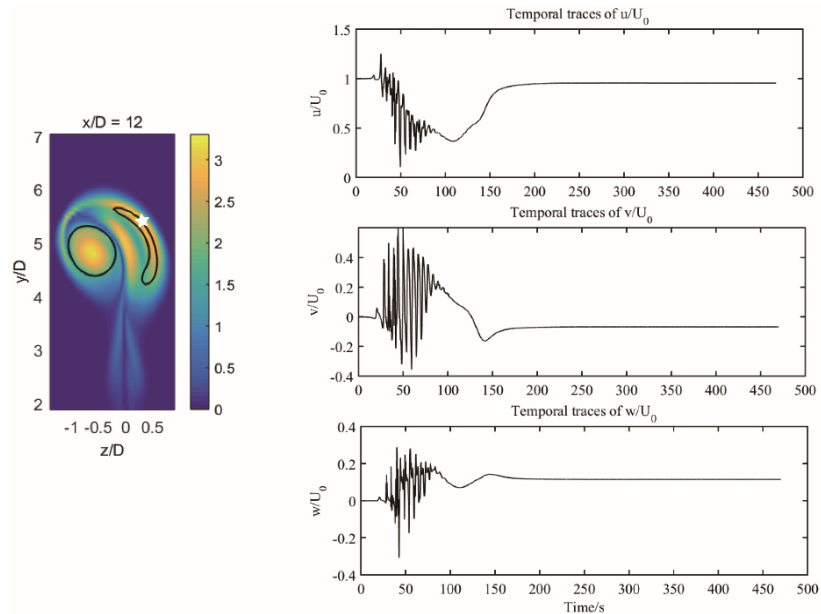

**Fig. 9** Time traces of three velocity components monitored at a sampling point in the $x/D = 12$ slice (3$^{rd}$ panel in Fig. 8). The sampling point is in the right-side vortex region and marked by a white star. $Re_D = 1000$.

It will be of interest to explore the degree of steadiness of the wake at $Re_D$ = 1000 in some further detail. Unsteadiness, if any, was therefore examined by means of time-traces of the velocity components measured at sampling points inside the vortex core regions. Fig. 9 and Fig. 10 show velocity histories monitored at $x/D$ = 12 and 24, respectively. The time histories in Fig. 9 show that the flow in the wake becomes completely independent of time after a transient stage has been passed. In spite of the major distortion of the vortex pair, seen by the highly asymmetric vorticity contours, the wake is undoubtedly still in a steady state $12D$ downstream. Similar results were shown at $x/D$ = 4 by Jiang *et al.* (2014).

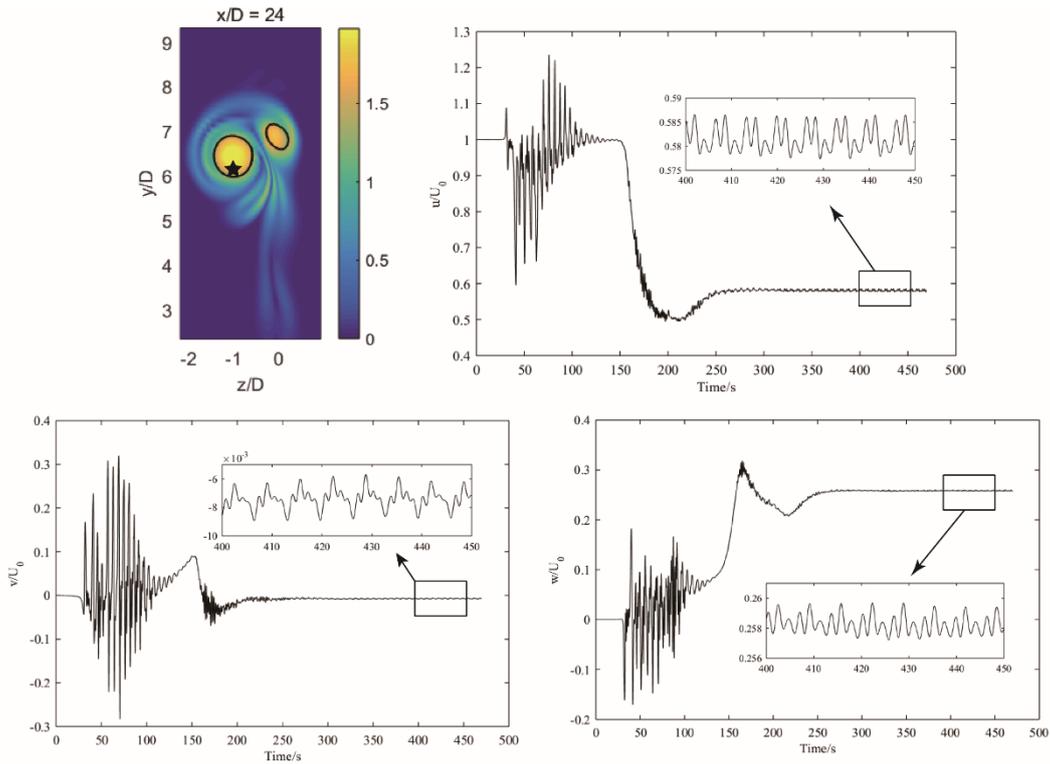

**Fig. 10** Same as Fig. 9, but at $x/D$ = 24 (4$^{th}$ panel in Fig. 8). The sampling point is in the left-side vortex core and arked by a black star. Zoom-in views are included to enable observations of minor unsteadiness. $Re_D$ = 1000.

Further downstream in the wake we observe from the corresponding results in Fig. 10 that the transient period is longer and the wake is more vigorus at $x/D$ = 24. After the transient period, all three velocity components settle to an apparently steady state. However, minute but non-negligible time-variations are seen in the insets. The time variations of the velocity components are apparently periodic but with an amplitude smaller than 1% of the free-stream velocity. The apparent periodicity makes us believe that the unsteadiness is a physical phenomenon rather than a numerical artifact. These observations suggest that the asymmetric wake behind the 45º inclined spheroid at $Re_D$ = 1000 is on the verge of becoming unsteady and that the unsteadiness develops downstream of $x/D$ = 14.

## 4.2 Details on wake transition and loss of symmetry

To further explore the destabilization process of the wake, two additional simulations were performed, one at $Re_D = 800$ and the other at $Re_D = 1200$. Snapshots of the resulting wakes are shown at the top and at the bottom of Fig. 11, respectively, to compare with the wake at $Re_D = 1000$ in the middle.

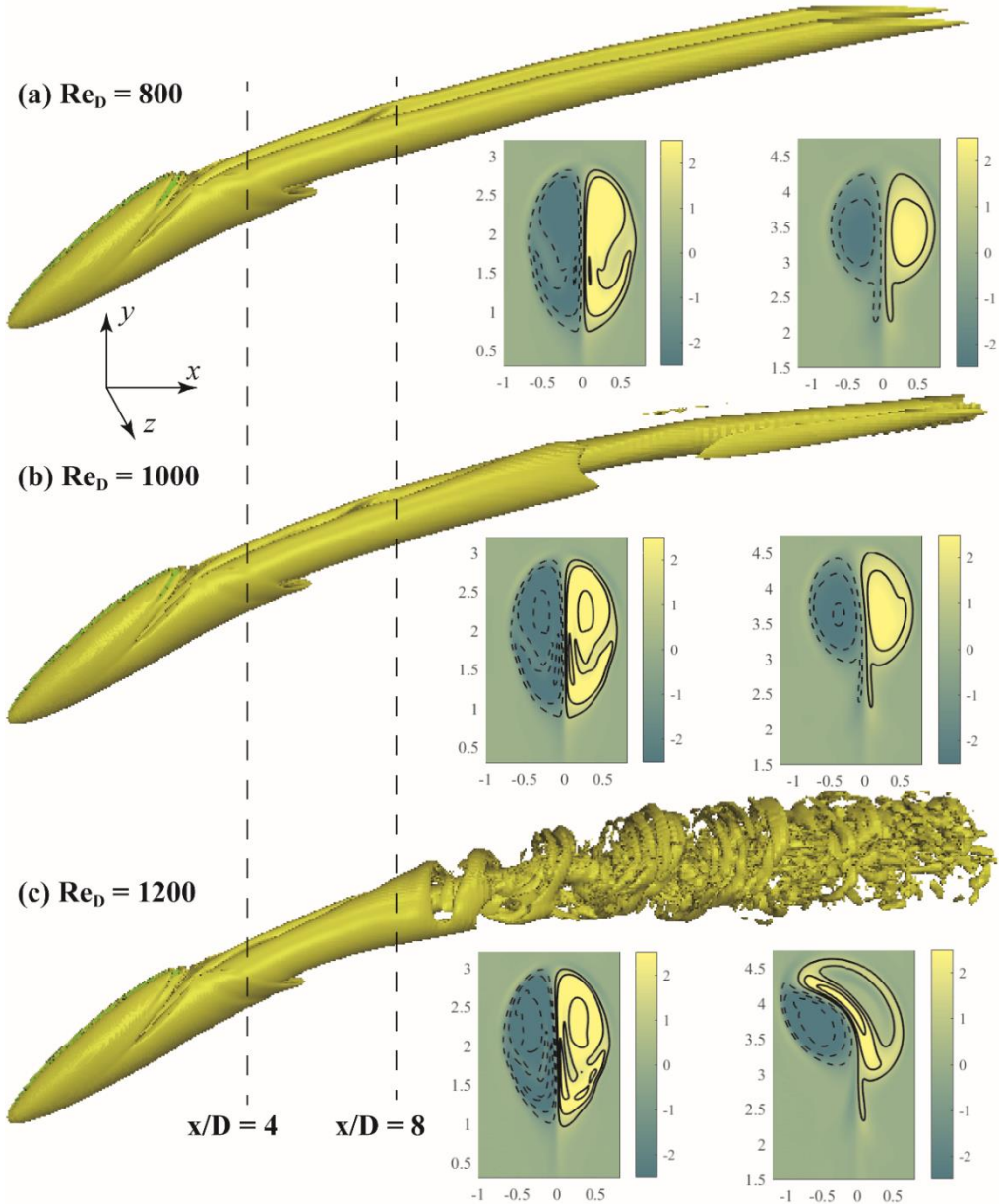

**Fig. 11** Comparisons of the wake at three different $Re_D$. In each subplot, the wake overall structure is shown by the iso-surfaces of $\lambda_2 = -0.5$. The distribution of streamwise vorticity $\langle \omega_x \rangle D/U_0$ in vertical slices at $x/D = 4$ and 8 are also shown. The contour lines in the $\langle \omega_x \rangle D/U_0$ colour plots are $\langle \omega_x \rangle D/U_0 = \pm 1, \pm 2,$ and $\pm 3$. Solid lines for positive values and dashed lines for negative values. (a) $Re_D = 800$; (b) $Re_D = 1000$; (c) $Re_D = 1200$.

The wake at $Re_D = 800$ is perfectly symmetric all the way throughout the computational domain, whereas the $Re_D = 1000$ and 1200 wakes are partially asymmetric. At $Re_D = 1200$ even the intermediate wake is highly asymmetric and significantly more complex than that at $Re_D = 1000$. However, the near-wake is still symmetric, similar as the wakes at $Re_D = 800$ and 1000. This is also shown by the force coefficient $C_{Fz}$ and the torque coefficients $C_{Mx}$ and $C_{My}$ in table 2, which are all of order $\sim 10^{-4}$ or $10^{-5}$. These coefficients are sufficiently close to zero so that their small values can be considered as a result of insufficient sampling in time. Since any asymmetries in the near-wake will make these coefficients non-zero, we conclude that the near-wake at $Re_D = 1200$ is symmetric despite its downstream development.

The time-averaged streamwise vorticity $<\omega_x>D/U_0$ distributions are shown in the vertical planes at $x/D = 4$ and 8 for all three $Re_D$ in Fig. 11. Both plots at $Re_D = 800$ are symmetric, whereas the plot at $x/D = 4$ is symmetric and that at $x/D = 8$ is slightly asymmetric at $Re_D = 1000$. At $Re_D = 1200$, the wake is heavily distorted and highly asymmetric at $x/D = 8$ even though the vortex pair remains almost symmetric at $x/D = 4$. We therefore conclude that although the wakes at $Re_D = 1000$ and $Re_D = 1200$ both are partially asymmetric, the inception of the asymmetry occurs more upstream as the Reynolds number increases, *i.e.* at $Re_D = 1200$.

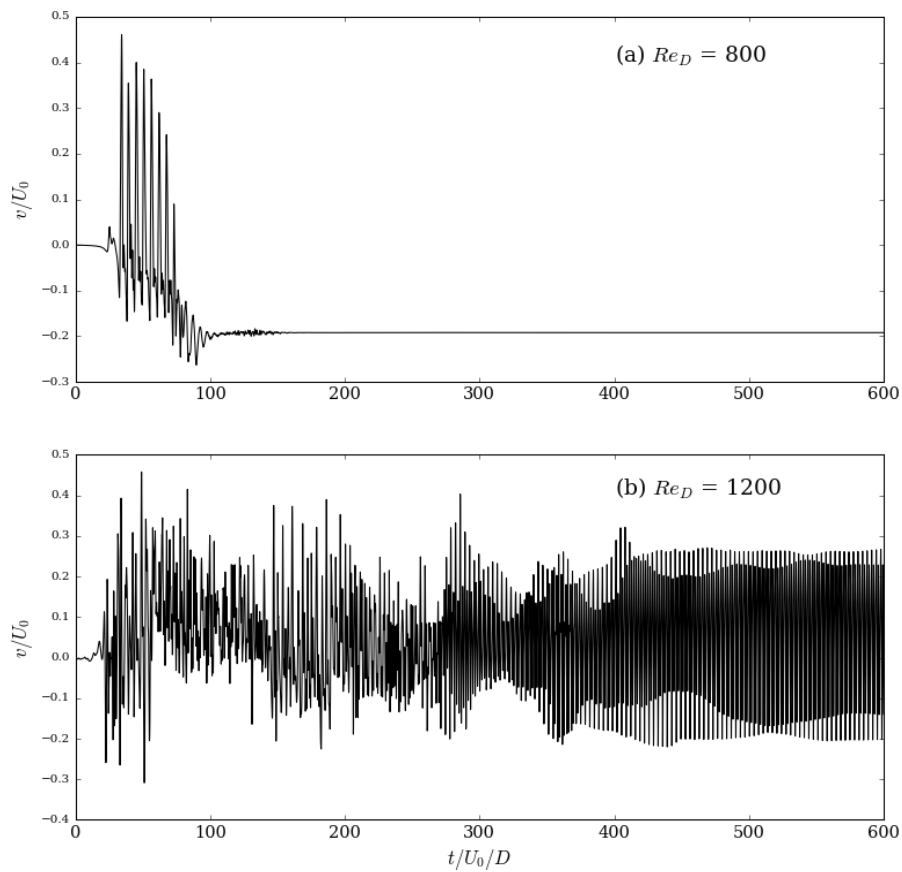

**Fig. 12** Time traces of the dimensionless vertical velocity $v/U_0$ at a sampling point at (18, 6, -1). (a) $Re_D = 800$, (b) $Re_D = 1200$.

The time dependency of the wakes is further examined in Fig. 12, where the time history of the vertical velocity compoent $v$ at a sampling point $18D$ downstream in the wake is shown for both $Re_D = 800$ and $1200$. The sampling location was chosen $4D$ downstream of $x/D = 14$ where we have learnt from Section 4.1 that minute unsteadiness exists in the $Re_D = 1000$ wake. We observe in Fig. 12(a) that even at $x/D = 18$, the $Re_D = 800$ wake is completely stationary after an early time-varying transient, whereas the velocity signal at $Re_D = 1200$ exhibits very strong oscillations throughout the simulation, which means the wake is already very unsteady. By recalling the time histories in Fig. 10, we can now conclude that the onset of unsteadiness in the wake occurs spontaneously somewhere in between $Re_D = 800$ and $1200$. This gives further evidence to our view that the wake at $Re_D = 1000$ is just on the verge of becoming unsteady and that the first inception of wake instabilities occurs at a $Re_D$ around 1000.

Fig. 13 shows the energy spectra for each of the three velocity components at a sample point in the $Re_D = 1200$ wake. The Strouhal number $St = fD/U_0 = 0.3$ corresponds to the primary frequency which is universal in the wake. In absence of vortex shedding, we anticipate that the primary frequency stems from the unsteadiness of the strain field generated by the stronger vortex and the induced rotation.

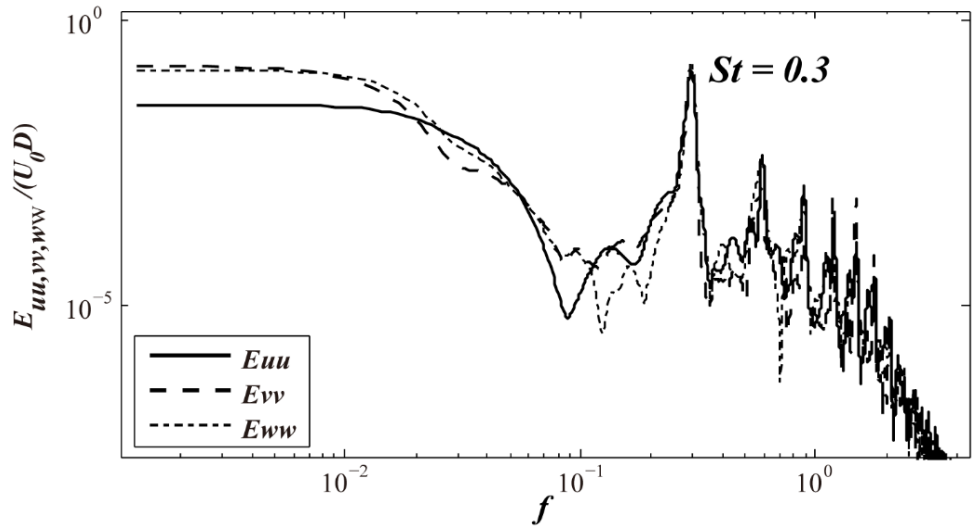

**Fig. 13** Energy spectra of the three velocity components at the sampling point (10, 4.5, -0.5). $Re_D = 1200$.

The interactions between the two vortices at $Re_D = 1000$ were discussed in section 4.1, suggesting that when the primary instability emerges, the strain field generated by the stronger vortex will rotate around itself at a certain frequency and make the weaker vortex to wrap around it and rotate. This scenario can now be proved by Fig. 14 where carefully selected snapshots of the instantaneous streamwise vorticity $\omega_x D/U_0$ are shown at $x/D = 11$. The time variation of $\omega_x$ was recorded for one period $T$, which was estimated from the primary Strouhal frequency $St = 0.3$. Four characteristic snapshots are presented, of which the first ($t = t_0$) and the last ($t = t_0 + T$) are indistinguisable, as shown in Fig. 14(a). This confirms beyond any doubt that the primary frequency is caused by the rotational motion displayed in Fig. 14. We notice that the stronger anticlockwise vortex A (depicted in Fig. 14 by dashed lines, *i.e.* $\omega_x < 0$) has a fairly fixed location in all snapshots, whereas the weaker vortex ($\omega_x > 0$ and depicted by solid contour lines) is severely deformed as it rotates around the stronger vortex.

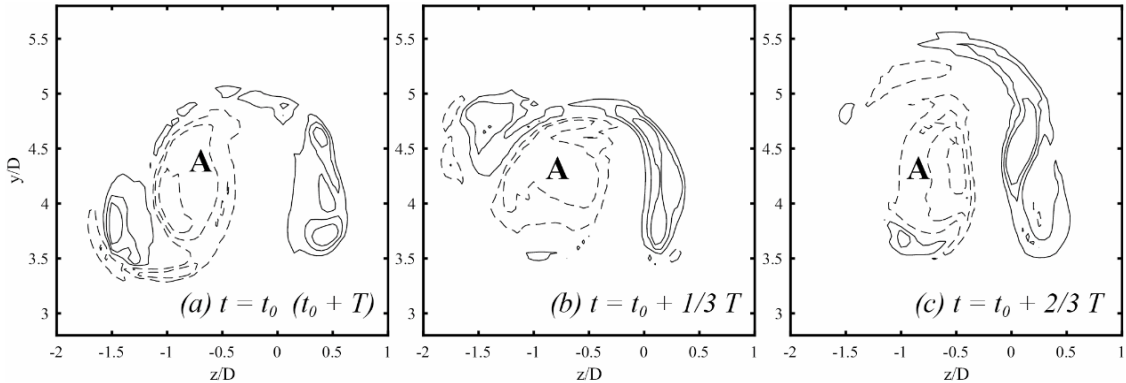

**Fig. 14** Snapshots of instantaneous streamwise vorticity $\omega_x D/U_0$ at $x/D = 11$ and at $Re_D = 1200$. The contour lines in all subplots are $\omega_x D/U_0 = \pm 1$, $\pm 2$, and $\pm 3$. Solid and dashed contour lines denote positive and negative values, respectively. (a) $t = t_0$ and $t_0 + T$, (b) $t = t_0 + 1/3T$, (c) $t = t_0 + 2/3T$, where $t_0 = 520\ D/U_0$.

## 5. Instability and Coherent Structures in the $Re_D = 3000$ Wake

As the Reynolds number is further increased to $Re_D = 3000$, the wake becomes more turbulent than at $Re_D = 1200$, and the instantaneous flow field is distinctly asymmetric, as shown in the overview in Fig. 15. Coherent laminar-like vortical structures are no longer visible in an instantaneous snapshot, but a persistent flow structure in the form of a concentrated vortex tube is pointed out in Fig. 15. In this section, we will first introduce the main features of the wake at $Re_D = 3000$ and thereafter report an in-depth study of the concentrated vortex structure. A grid convergence test was presented by Jiang *et al.* (2015b) and will not be repeated here.

### 5.1 The asymmetric turbulent wake at $Re_D = 3000$

Although the primary characteristics of the laminar wake at $Re_D \sim 10^3$ in Fig. 11, *i.e.* the counter-rotating vortex pair, is no longer observable in the instantaneous flow field in Fig. 15, a vortex pair can nevertheless be observed in time-averaged flow fields. The time-averaged streamwise vorticity $<\omega_x>D/U_0$ distribution over vertical planes at *x/D* = 6 and 8 is shown in Fig. 16, from which the existence of a counter-rotating vortex pair is evident. However, the vortex pair is distinctly asymmetric with the almost circular right-side vortex being more concentrated and having larger vorticity, while the left-side vortex is deformed and has lower vorticity. The asymmetric vortex pair is shifted in the positive *z*-direction, *i.e.* towards the right with respect to the geometrical symmetry plane at *z/D* = 0. It is interesting to recall that the wakes at all lower Reynolds numbers were deflected to the left. However, this is not a Reynolds number effect. The wake will arbitrarily be deflected to one side, and this ambiguity stems from a pitchfork bifurcation. The appearance of either right-side or left-side deflections offers evidence of this conjecture.

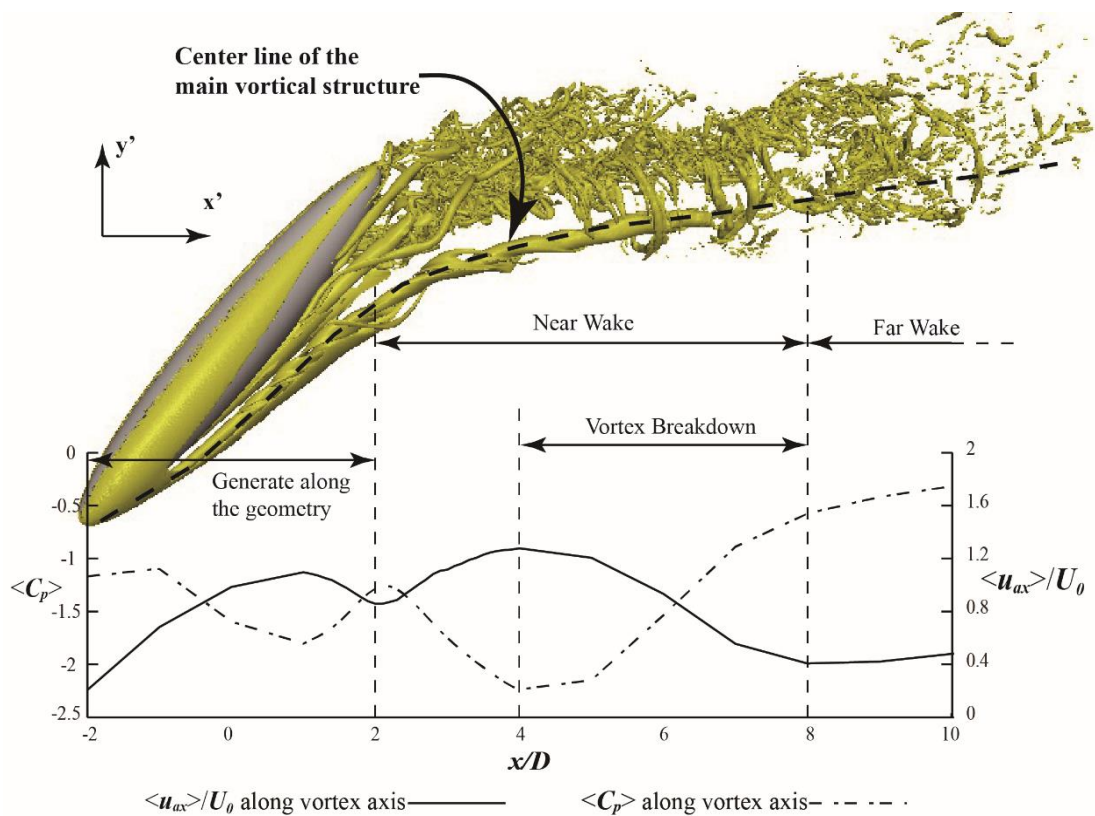

**Fig. 15** Overview of the wake at $Re_D = 3000$, visualized by an instantaneous iso-surface of $\lambda_2 = -20$. The centerline of the coherent tube-like structure is indicated. The variation of the time-averaged axial velocity $<u_{ax}>/U_0$ and the mean pressure coefficient $<C_p> = (<p> - p_0)/0.5\rho U_0^2$ (where $p_0$ is the reference pressure at the inlet) along the centerline are plotted down to *x/D* = 10. (Reprinted from Jiang, Gallardo, Andersson, Okulov. On the peculiar structure of a helical wake vortex behind an inclined prolate spheroid. *J. Fluid Mech.* 801, 1-12, 2016, with the permission of Cambridge University Press.)

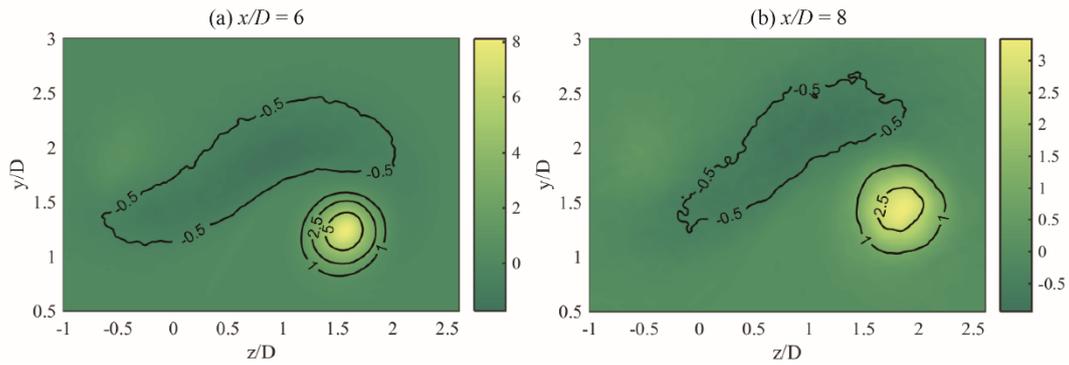

**Fig. 16** Time-averaged streamwise vorticity $<\omega_x>D/U_0$ distribution over two x-slices at $Re_D = 3000$. (a) $x/D = 6$; (b) $x/D = 8$.

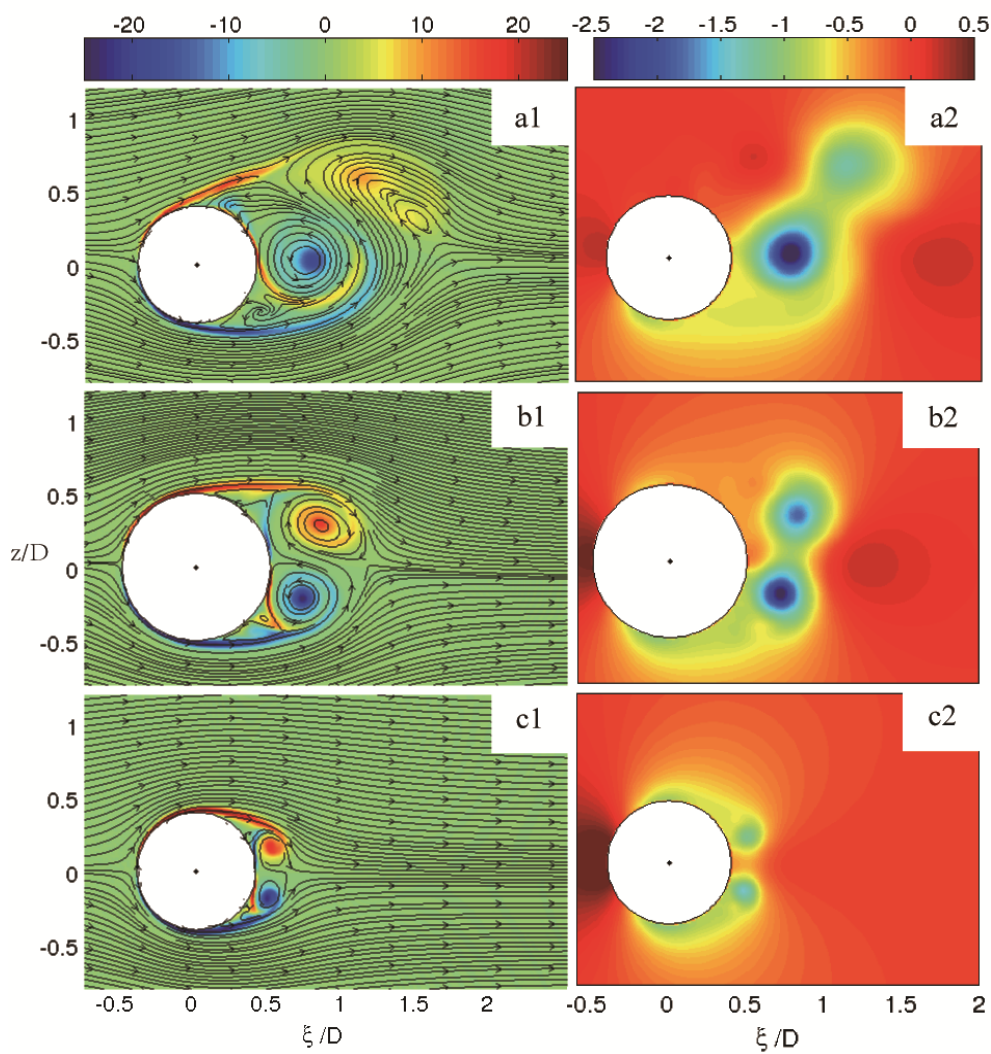

**Fig. 17** Left column: time-averaged axial vorticity $<\omega_\eta>D/U_0$ superimposed on streamlines of mean velocity vectors projected into the plane. Right column: the mean pressure coefficient distribution. All plots are in the fixed-to-body coordinate system ($\xi, \eta, z$) defined in Fig. 1 and are viewed in $+\eta$ direction. (a) $\eta/D = +1.8$, (b) $\eta/D = 0$, (c) $\eta/D = -1.8$. $Re_D = 3000$. (Reprinted from Jiang, Gallardo, Andersson, Zhang. The transitional wake behind an inclined prolate spheroid. *Phys. Fluids.* 27, 093602, 2015, with the permission of AIP Publishing.)

The asymmetric mean flow field at $Re_D = 3000$ can also be seen from the side-force coefficient $C_{Fz}$ in table 2. Different from the lower $Re$ cases, where the side-forces were essentially zero, thus indicating a perfectly symmetric near-wake, the side-force (mean value) at $Re_D = 3000$ is surprisingly large. In fact, we notice that the side force amounts to about 75% of the streamwise force $F_x$, *i.e.* almost comparable to the drag force on the spheroid. This stems from a severely distorted near-wake, which is distinctly different from all the lower Reynolds number cases. A closer look at the asymmetric near-wake is provided in Fig. 17, where the flow field around the spheroid is presented in the fixed-to-body coordinate system ($\xi, \eta, z$) defined in Fig. 1. The mean axial vorticity $<\omega_\eta> D/U_0$ and the mean pressure field are plotted in Fig. 17, from which we observe that a modest asymmetry in the wake is visible even close to the lower pole (Fig. 17c), *i.e.* when the vortex sheets have just separated from the spheroid and begun to roll up. The pressure field is accordingly asymmetric almost all over the surface of the spheroid from the lower pole to the upper pole, and the asymmetry becomes more and more pronounced towards the upper pole (trailing tip). The asymmetric pressure distribution seen in Fig. 17 results in a strong pressure force in the negative z-direction, *i.e.* consistent with $C_{Fz} < 0$ in table 2. It is therefore evident that the strongly asymmetric time-averaged pressure field gives rise to the strong side force.

The substantial side-force together with the highly asymmetric time-averaged flow field lead us to some important conclusions. Firstly, we can infer that the instantaneous wake remains asymmetric and deflected in the positive z-direction. In other words, no periodic or quasi-periodic flipping of the wake from one side to the other occurs, and no alternating vortex shedding takes place from the inclined spheroids, at least not at this $Re$. Secondly, similar strong asymmetric vortex wakes were supposed to occur behind sharp-nosed slender bodies (*e.g.* fighter forebodies, conical-cylinder bodies) only at high attack angles (Zeiger *et al.* 2004). The aspect ratios of such slender bodies are typically larger than 10:1. In the marine engineering field, where a 6:1 prolate spheroid is commonly used as simplified submarine or other underwater vehicle's hull shape, a wake symmetry assumption is widely used, both for numerical and experimental studies, see *e.g.* Gross *et al.* (2011). This is why Ashok and Smits (2013) and Ashok *et al.* (2015) claimed their observations of an asymmetric wake in their submarine model experiment to be "surprising". Although submarine model tests were carried out at much higher Reynolds numbers ($Re_L \sim 10^6$) and somewhat lower attack angles (lower than 15°), the physical mechanism may be similar as in the present study. There is no doubt that these observations will bring new insight in the maneuvering of marine vehicles.

A long-lasting debate about the instability mechanism in the sharp-nosed slender-body wakes (Bridges 2006) has been concerned about whether the instability is of convective or global nature. The debate has not yet been concluded, but there seems to be an inclination towards the convective instability because one must manually introduce a perturbation on the model in order to obtain a time-averaged asymmetric wake in sharp-nosed slender-body studies, regardless whether it is a numerical or experimental study. In the present study, however, no artificial asymmetric disturbances were introduced (the computational domain, the spheroidal geometry, all boundary conditions *etc.* were symmetric). On the other hand, the prolate spheroid is blunter than sharp-nosed slender bodies, and it has been widely accepted that a blunt nose is much less sensitive to disturbances than sharp noses (Moskowitz *et al.* 1989). We are therefore inclined to attribute the deflection of the wake behind the 45º inclined spheroid to a global instability. This conjecture is far from being a solid conclusion.

**5.2 The main coherent structure in the wake**

In this subsection, we aim to explore the exceptionally long and persistent coherent vortical structure in the $Re_D = 3000$ wake, clearly seen in the instantaneous view presented in Fig. 15. The circular region of concentrated positive vorticity in Fig. 16 is a manifestation of the persistency of the tubular vortex structure. It is somewhat surprising that such a significant coherent structure can be so persistent in this complex and otherwise unsteady wake at a relatively high Reynolds number. To better understand this vortical structure, we first examine the relationship between the velocity and vorticity field inside the vortex. This can be achieved by means of the helicity density $H_d = \vec{V} \cdot \vec{\omega}$, which was originally defined by Moffatt (1969) and later re-introduced as a means of graphical representation of three-dimensional flow fields by Levy et al. (1990). The helicity density essentially plays the role as a filter, which filters out regions with either low vorticity or low velocity or regions in which the local velocity vector is at a large angle with the vorticity vector. Since the main vortex tube is a persistent structure in the otherwise unsteady wake, it is appropriate to study the time-averaged flow field. We therefore introduce the approximate mean helicity density defined as:

$$\widetilde{H} = D/U_0^2 \left( \sum_{i=1}^{3} <u_i><\omega_i> \right), \ i = x, y, z \tag{6}$$

We refer to $\widetilde{H}$ as "approximate" mean helicity density because it differs from the exact time-averaged helicity density $<H_d>$ by a term $<u_i'\omega_i'>$, where $u_i'$ and $\omega_i'$ represent fluctuating parts of the velocity and vorticity components. The neglected term $<u_i'\omega_i'>$ is the projection of fluctuating vorticity vector on the fluctuating velocity vector. Since this main coherent structure is a concentrated and persistent vortex tube, we have good reasons to believe that these fluctuating quantities are modest. Accordingly, $\widetilde{H}$ serves as a reliable approximation of the mean helicity density.

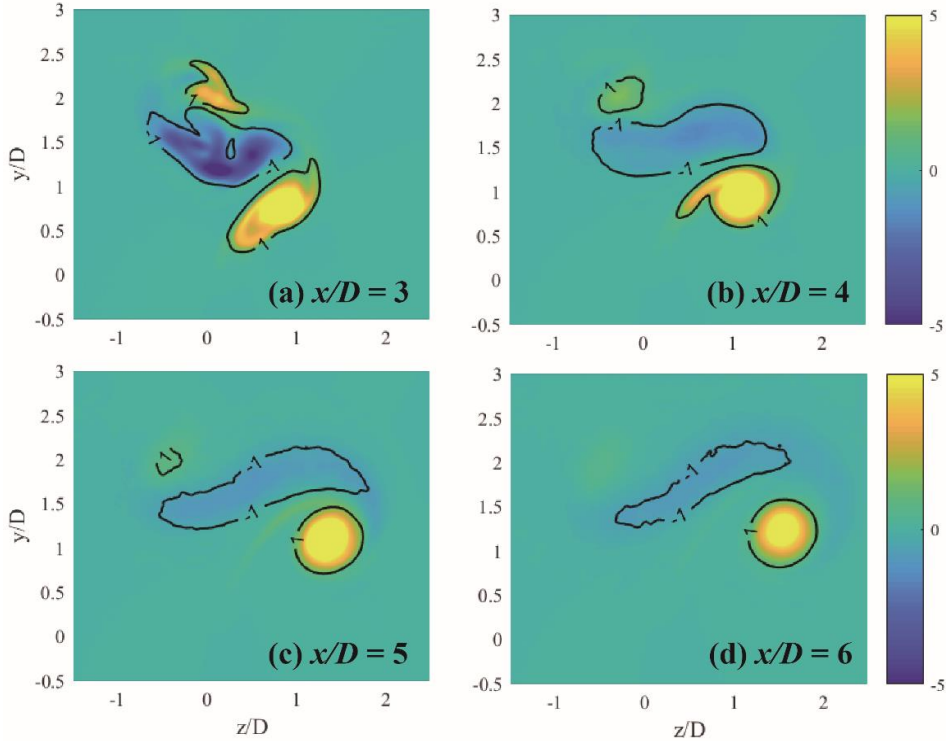

**Fig. 18** Time-averaged approximate helicity density $\widetilde{H} = D/U_0^2 \, (\sum_{i=1}^{3} <u_i><\omega_i>)$ distribution at four different streamwise locations in the near-wake. The contour lines represent $\widetilde{H} = \pm 1$. (a) $x/D = 3$; (b) $x/D = 4$; (c) $x/D = 5$; (d) $x/D = 6$. $Re_D = 3000$.

The distribution of $\widetilde{H}$ in four different planes is shown in Fig. 18, from which we realize that the helicity inside the coherent vortex tube is very strong. Inside the coherent vortex tube, we have $\widetilde{H} > 0$, which means that the directions of the velocity vector and the vorticity vector coincide. In the regoin associated with the weaker vortex region, we observe negative $\widetilde{H}$-values, which means that the vorticity vector is in the opposite direction of the velocity vector. The close resemblance between the plots in Fig. 18(d) and Fig. 16(a), both at the same *x*-location, implies that $<\omega_x>$ is the dominant component of the vorticity vector in both vortex regions, and, furthermore, that the mean streamwise velocity $<u>$ is the dominant velocity component in these regions. The $\widetilde{H}$-contours in Fig. 18 show beyond any doubt that the main coherent structure in the $Re_D = 3000$ wake is a concentrated helical vortex (Alekseenko *et al.* 1999; 2007).

One can observe from Fig. 15 that the dashed vortex centerline is bent, whereas Fig. 16 showed that the vortex tube is deflected towards the right. This coherent vortex tube is thus a complex three-dimensional flow structure. Neither the global coordinate system (*x, y, z*) nor the fixed-to-body coordinate system (*ξ, η, z*) is appropriate for detailed examinations of the helical vortex tube. To serve this purpose a local cylindrical coordinate system, whose axial direction exactly follows the three-dimensional curved vortex centerline, is used. The vortex centerline is defined as the locus of local pressure minima in the vortex tube at different *x* locations. A velocity vector in the Cartesian coordinate system can now be transformed to the local cylindrical coordinate system and written as $\vec{u} = (u_{ax}, u_\theta, u_r)$, where the three components denote the local axial velocity, azimuthal velocity, and radial velocity, respectively. The vorticity vector is transformed in the same way.

The time-averaged axial velocity $<u_{ax}>$ and the pressure coefficient $<C_p>$ along the vortex centerline are plotted in the lower part of Fig. 15. The axial velocity increases gradually from the lower tip ($x/D \approx$ -2) up to $x/D \approx 1$ where it peaks at about $1.1U_0$, *i.e.* 10% larger than the free-stream velocity. This high velocity indicates a strong axial flow along the leeward side in the near wake and originates from the vorticity separated from the inclined separation line on the surface of the spheroid. Then, as the vortex tube deflects to the right side and moves out of the shadow of the spheroid to meet the free-stream, the axial velocity is slightly reduced while the pressure coefficient increases. The slight weakening of the vortex tube at around $x/D$ = 2 stems from the sudden influence of the free-stream. However, the helical vortex recovers and persists its coherence down into the intermediate wake at $x/D \approx 6 - 7$. The visualized wake in Fig. 15 suggests that the vortex tube somehow breaks up at around $x/D = 6$, downstream of which the vortex tube can only be observed in the time-averaged flow field. This break-up phenomenon is related to a helical symmetry alteration which will be discussed in the next subsection. Before that, let us proceed with an interpretation of the results in Fig. 15.

A striking observation is the opposite trends of the variations of $<u_{ax}>$ and $<C_p>$ in Fig. 15. Indeed, the sum $<C_p> + (<u_{ax}>/U_0)^2$ turns out to remain almost constant from $x/D$ = -1 to $x/D$ = 7. If this apparently constant sum is multiplied by $0.5\rho U_0^2$, we get the following equation:

$$<p> + 0.5\rho <u_{ax}>^2 \approx const. \qquad (7)$$

Considering the relatively high Reynolds number and the complexity of the wake, the neat relationship in equation (7) is no doubt surprising. Moreover, equation (7) resembles the mechanical energy conservation law, namely the Bernoulli principle, which says that the total mechanical energy is constant along a streamline in an incompressible inviscid flow. The only difference between the Bernoulli principle and equation (7) is that also potential energy is accounted for in the Bernoulli principle. In the present study, however, the gravity force was neglected from the very beginning, which means that potential energy considerations are irrelevant herein. Therefore, to justify the apparent validity of equation (7) in the present case, the underlying assumptions of "inviscid" flow "along a streamline" should be considered.

Firstly, we can easily estimate from Fig. 16 that the diameter of the vortex tube is approximately the same as *D*, and from Fig. 15 that the axial velocity is about the same as $U_0$ in the wake region from $x/D$ = -1 to $x/D$ = 7. Hence a local Reynolds number in the vortex tube can be estimated as $Re_{tube} \sim U_0D/\upsilon = Re_D = 3000 >> 1$. This suggests that viscous effects are negligible inside the vortex tube. Secondly, we calculated both $0.5\rho <u_{ax}>^2$ and $0.5\rho(<u>^2+<v>^2+<w>^2)$ along the vortex centerline in this wake region. The close comparison showed that the kinetic energy associated with $<u_{ax}>$ accounted for more than 99% of the total kinetic energy. In other words, the other two velocity components $u_\theta$ and $u_r$ together contribute less than 1% of the kinetic energy along the vortex centerline and are therefore deemed negligible. Thus, although the centerline of the vortex tube was identified as the locus of local pressure minima, it can also be safely approximated as a streamline. The validity of the surprisingly simple energy conservation law (7) has therefore been justified.

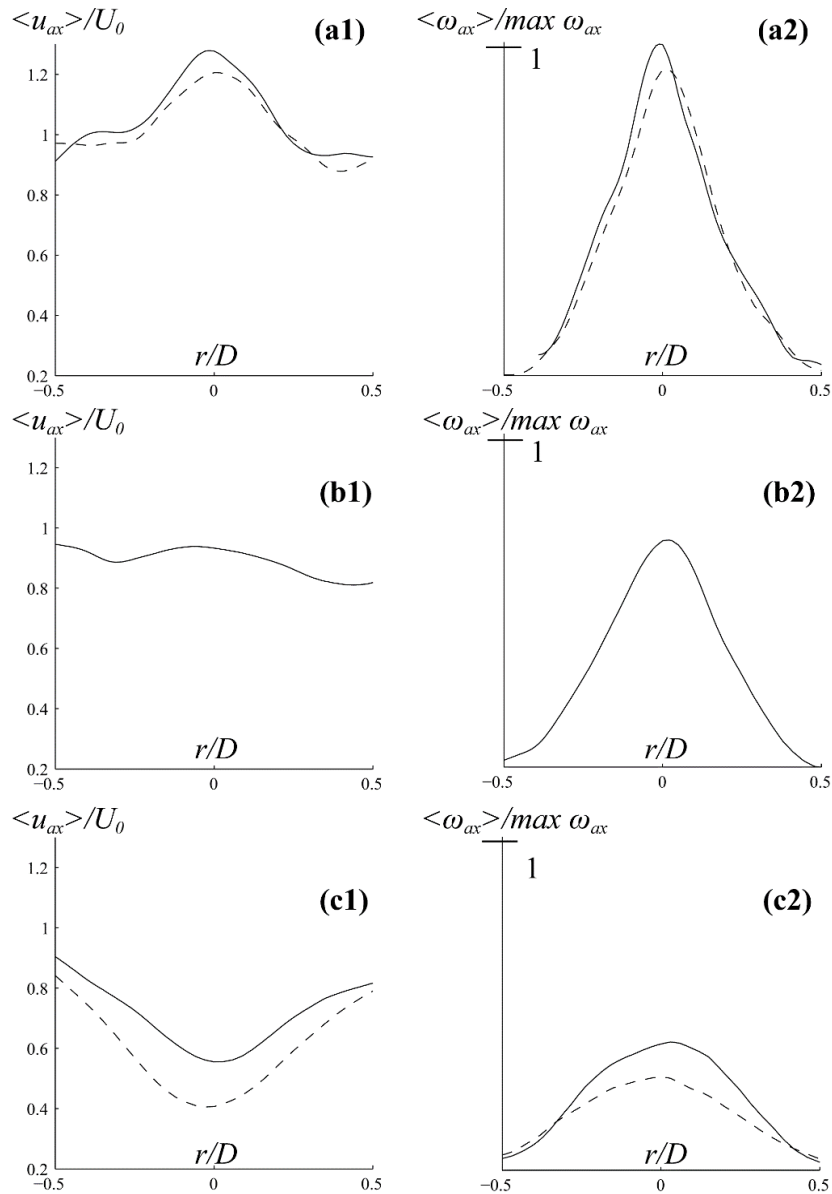

**Fig. 19** Profiles of axial velocity $<u_{ax}>/U_0$ (left column) and axial vorticity $<\omega_{ax}>$ (right column) along a diameter of the helical vortex cross-sections. The results are from five different cross-sections whose centers locate at (a) $x/D = 4$ (solid) and 5 (dashed); (b) $x/D = 6$; and (c) $x/D = 7$ (solid) and 8 (dashed), respectively. $Re_D = 3000$.

**5.3 A new helical symmetry alteration scenario**

To facilitate the further discussions, we divide the main coherent structure in the $Re_D$ = 3000 wake, *i.e.* the helical vortex tube, into three parts (or stages), as indicated in Fig. 15. The first stage -2 < $x/D$ < +2 is referred to as the *generation stage*, the second stage +2 < $x/D$ < +8 is referred to as the *near-wake*, in which we will see a new helical symmetry alteration scenario, and the third stage beyond $x/D$ = +8 is referred to as the *far wake*, in which self-similarity will be examined.

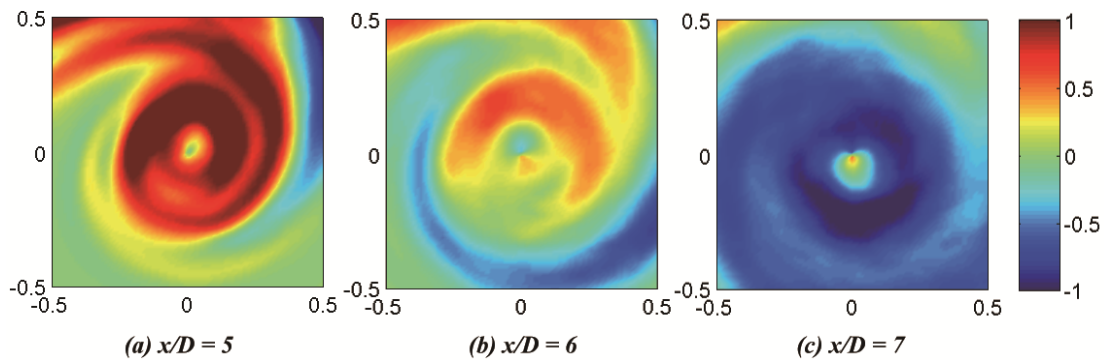

**Fig. 20** Azimuthal vorticity $<\omega_\theta>D/U_0$ distribution in three different vortex tube cross-sections whose centers locate at (a) $x/D$ = 5; (b) $x/D$ =6; and (c) $x/D$ = 7. (Reprinted from Jiang, Gallardo, Andersson, Okulov. On the peculiar structure of a helical wake vortex behind an inclined prolate spheroid. *J. Fluid Mech.* 801, 1-12, 2016, with the permission of Cambridge University Press.)

Although the wake at $Re_D$ = 3000 is more unsteady and asymmtric than at lower Reynolds numbers, the primary features of the generation stage are still the separation and rolling-up of vortex sheets. As the Reynolds number increases, the separated vortex sheets can no longer maintain a smooth surface. Yet the mechanism is the same as we have explained before. It is worth to mention that the helical motion has already been established in this generation stage, when vorticity components are produced in the three-dimensional flow separation process, and the axial velocity is induced by these vorticity components.

The newly generated vortex tube has initially an inclination angle close to 45º, but is soon bent and tilted as it enters the near-wake stage. In the near-wake stage, the vortex topology changes dramatically inside the helical vortex tube from $x/D$ = 4 to $x/D$ = 8, as indicated in Fig. 15.

In Fig. 19, the variations of both axial velocity $<u_{ax}>$ and axial vorticity $<\omega_{ax}>$ are shown over some different cross-sections of the helical vortex. In the right column of Fig. 19, we notice that despite the decreasing peak values, the shape of the $<\omega_{ax}>$-profiles remain the same with maximum axial vorticity always at the center of the vortex tube. On the contrary, the profiles of the axial velocity $<u_{ax}>$ exhibit distinct topological changes.

Unexpected jet-like axial velocity profiles at $x/D = 4$ and 5 are observed in Fig. 19(a1). Such jet-like velocity profiles are not commonly seen in wakes, although similar profiles were recently reported behind a flying bird or a swimming fish by Taylor *et al.* (2013) who concluded that such jet-like velocity profiles help to improve the efficiency of the movement of the bird and fish. The unforeseen velocity profiles in Fig. 19(a1) are most likely induced by the three-dimensional effects in the near wake.

Further downstream of $x/D = 5$, the axial velocity inside the helical vortex tube experiences a dramatic change from the jet-like profile to a wake-like profile, as clearly shown in Fig. 19(c1). Just in the transition between jet- and wake-like profiles, *i.e.* at $x/D = 6$ (Fig. 19(b1)), the axial velocity is almost uniform over the cross-section. The remarkable changes of axial velocity profile should be considered together with the results for the azimuthal vorticity $<\omega_\theta>D/U_0$ in Fig. 20. As shown in Fig. 20(a), the azimuthal vorticity is positive inside the vortex core when the axial velocity has a jet-like profile (Fig. 19(a1)). According to Fig. 20(c), however, the azimuthal vorticity has become negative inside the vortex core at $x/D = 7$ where the axial velocity profile in Fig. 19(c1) is jet-like. No clear-cut sign of $<\omega_\theta>$ can be inferred form Fig. 20(b) at $x/D = 6$, which corresponds to transition from jet-like to wake-like axial velocity. These results suggest that the sudden change of the shape of the axial velocity profile inside this helical vortex is closely related to the change-of-sign of the azimuthal vorticity. It is well-known that a helical vortex with positive azimuthal vorticity in the vortex core is a right-handed helical vortex, whereas a vortex with negative azimuthal vorticity inside the core makes it a left-handed helical vortex (Okulov 1996). The process observed in Fig. 19 and Fig. 20 therefore reveals a helical symmetry alteration from right to left (Martemianov and Okulov 2004).

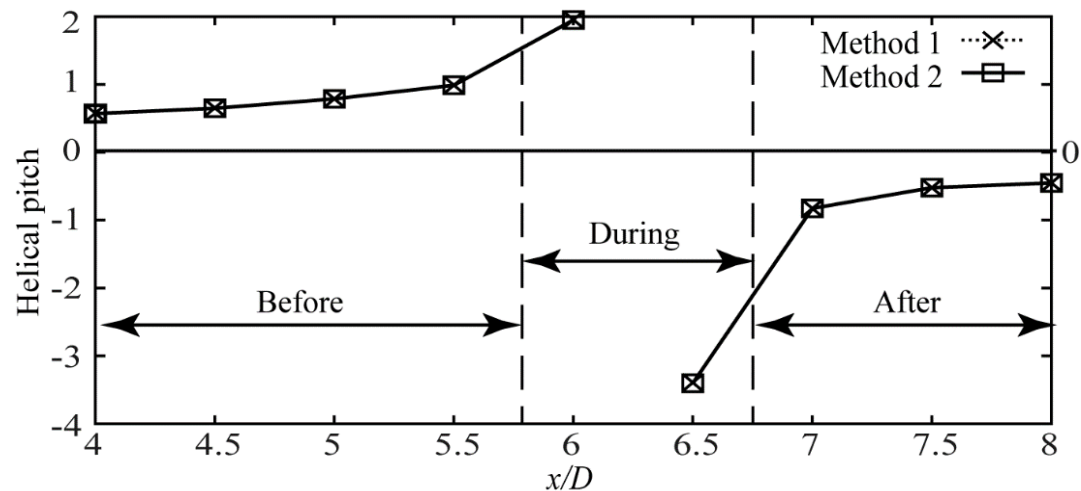

**Fig. 21** Variation of the helical pitch *l* results along the helical vortex centerline. Results calculated by means of two different methods are plotted for comparison. $Re_D = 3000$. (Reprinted from Jiang, Gallardo, Andersson, Okulov. On the peculiar structure of a helical wake vortex behind an inclined prolate spheroid. *J. Fluid Mech.* 801, 1-12, 2016, with the permission of Cambridge University Press.)

The alteration from a right-handed helical vortex to a left-handed one, accompanied by a change of axial velocity profile, are normally a strong indication of a vortex breakdown. However, as part of a "classical" vortex breakdown scenario, one must also be able to observe a flow reversal in the vortex center, as clarified in Leibovich (1978; 1984). From Fig. 19(a1) to (c1) we could not observe any clear evidence of a flow reversal; in fact, the axial velocity variation along the vortex axis, as sketched in Fig. 15, shows beyond any doubt that the axial velocity remains positive all through the transition. The possible existence of flow reversal essentially means that a recirculation bubble with a stagnation point can be observed. It should be mentioned that a similar phenomenon was observed by Sarpkaya (1971) and Faler and Leibovich (1977) in low swirl-level flows, but neither of these studies considered any helical symmetry alteration.

Martemianov and Okulov (2004) derived in a theoretical study several different scenarios that may occur associated with a helical symmetry alteration. These were characterized by abrupt changes of different velocity profiles as the azimuthal vorticity changed sign. The "classical" vortex breakdown is the most common one, where the helical symmetry change is accompanied by a switch from jet-like to wake-like velocity profile and the occurence of a recirculation bubble. Yet Martemianov and Okulov (2004) also hypothesized more possible scenarios in which a flow reversal does not exist. Velte *et al.* (2011) identified one of the hypothesized scenarios in an experiment, and named it as a "slight" helical symmetry alteration. The "slight" helical symmetry alteration is characterized by a transition between two wake-like profiles accompanied by a vortex core expansion. The process in the present study is, however, different from both the "classical" and the "slight" scenarios. The helical vortex in the present wake experiences a sudden change of axial velocity profile from jet-like to wake-like, during which the right-handed helix transforms to a left-handed one, but maintains the positive axial velocity all the way without any flow reversal. Although this scenario is one of the hypothesized scenarios that theoretically may happen (Okulov and Sørensen 2010), it has never been reported from any experimental or numerical study before.

Okulov (1996) first pointed out that the helical symmetry alteration can be identified quantitatively by a change-of-sign of the helical pitch *l*. Velte *et al.* (2009) used two different methods to calculate the helical pitch, which we also adopt in the present study. The first method uses a least-square method to minimize the residual of:

$$u_{ax} + \left(\frac{r}{l}\right)u_\theta = u_a(x) \tag{8}$$

where $u_a$ denotes the vortex advection velocity. The second method calculates the helical pitch *l* from the expression:

$$l = \frac{-\int_\Sigma \rho u_\theta u_{ax} r \, d\Sigma}{(\int_\Sigma \rho u_{ax}^2 \, d\Sigma - u_a G)} \quad (9)$$

where Σ denotes the cross-sectional area of the vortex and $G$ is the mass flow rate.

Results obtained with both methods are shown in Fig. 21, from which we immediately observe the collapse of the results obtained with the two methods. The helical pitch $l$ is positive before the helical alteration occurs at $x/D \approx 6$, while the helical pitch is negative downstream of the alteration, *i.e.* $x/D > 6$. The mangitude of the helical pitch increases rapidly near $x/D \approx 6$ and thus exhibits a singularity-like behaviour. This is an interesting observation in view of Okulov *et al.* (2005) who pointed out that when a helical symmetry alteration occurs, the helical pitch may either be zero (as acquired by a vortex ring) or go to infinity. The former of these alternatives is undoubtedly the most common one, whereas a pitch tending to infinity, as suggested by the results in Fig. 21, is rare.

The change-of-sign of the helical pitch $l$ proved that the helical vortex tube, *i.e.* the main coherent structure in the inclined prolate spheroid wake, experiences a helical symmetry alteration process. Therefore, a complete description of the helical vortex tube reads as follows: a right-handed helical vortex with positive azimuthal vorticity in the vortex core and a jet-like axial velocity profile experiences a sudden helical symmetry alteration into a left-handed helical vortex with negative azimuthal vorticity in the core and a wake-like velocity profile. Throughout the alteration process, no flow reversal can be observed along the vortex axis. Although Martemianov and Okulov (2004) more than a decade ago proved such a scenario to be possible, this is probably the first time that the scenario is detected and analyzed.

## 5.4 The self-similarity law in the far wake

In the far fake, *i.e.* downstream of $x/D \approx 8$, fairly concentrated and quasi-axisymmetric mean streamwise vorticity distributions can be observed all the way to the outlet of the computational domain. The coherent helical vortex has already experienced the helical symmetry alteration described in subsection 5.3 and therefore appears with positive axial vorticity, negative azimuthal vorticity, and wake-like axial velocity profiles. The quasi-axisymmetric wake distribution motivated us to examine the validity of the well-known self-similarity law proposed by George (1989) and Johansson *et al.* (2003). The self-similarity law is attractive for practical purposes because the main features of the wake are represented by simplified expressions. However, the self-similarity was originally based on the assumption of perfect axisymmetry and its universality has therefore been doubted. However, Okulov *et al.* (2015) recently reported an abnormal axisymmetric wake, in which the self-similarity law was unexpectedly satisfied. Fluid velocity variations are usually rather small in the far wake, which makes it a challenge to obtain sufficiently detailed and accurate data from laboratory experiments. DNS, on the other hand, has no such limitations and the computed flow field is accurate even in the far wake.

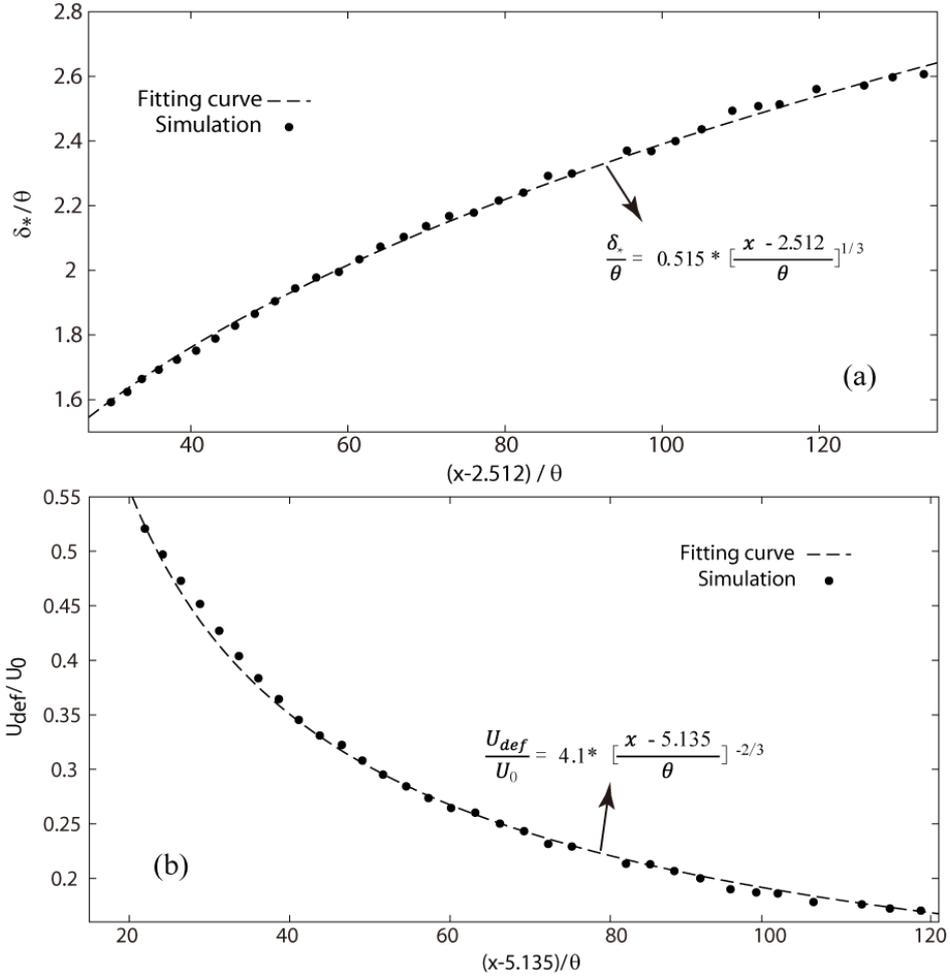

**Fig. 22** Validation of the self-similarity law in the far wake at $Re_D = 3000$. (a) $\delta_*/\theta$ vs. $(x - x_{01})/\theta$, where $x_{01} = 2.512$ ; (b) $U_{def}/U_0$ vs. $(x - x_{02})/\theta$, where $x_{02} = 5.135$. The scattered black dots are data obtained from the simulation, while the dashed lines are the best-fit curves (13) and (14).

The velocity deficit $U_{def} = (U_0 - u_{ax})_{max}$ and the length scale $\delta_*$ of the vortex along the axis should follow the self-similarity laws:

$$\frac{\delta_*}{\theta} = a\left[\frac{(x-x_0)}{\theta}\right]^{1/3} \tag{10}$$

$$\frac{U_{def}}{U_0} = b\left[\frac{(x-x_0)}{\theta}\right]^{-2/3} \tag{11}$$

where the momentum thickness $\theta$ and transverse length $\delta_*$ of the vortex are defined as:

$$\theta^2 = \lim_{\tilde{r}\to\infty} \frac{1}{U_0^2} \int_0^{\tilde{r}} u_z(U_0 - u_z)r\,dr \;;\; \delta_*^2 = \lim_{\tilde{r}\to\infty} \frac{1}{U_{def}} \int_0^{\tilde{r}} (U_0 - u_z)r\,dr. \qquad (12)$$

The self-similarity laws (10) and (11), see e.g. Johansson *et al.* (2003), imply that the maximum velocity deficit along the vortex axis in the far wake decreases as $x^{-2/3}$, whereas the width of the vortex grows as $x^{1/3}$.

Here, far-wake data from $x/D = 9$ to 26 have been deduced form the present DNS and plotted in Fig. 22. The best-fit curves based on DNS data:

$$\frac{\delta_*}{\theta} = 0.515 * \left[\frac{x - 2.512}{\theta}\right]^{1/3} \qquad (13)$$

$$\frac{U_{def}}{U_0} = 4.1 * \left[\frac{x - 5.135}{\theta}\right]^{-2/3} \qquad (14)$$

are also included. The comparisons show that the fitted self-similarity laws (13) and (14) match surprisingly well with the data from this complex and asymmetric wake behind the 45º inclined 6:1 prolate spheroid.

Moreover, the numerical values $a = 0.515$ and $b = 4.1$ estimated from the DNS data give $a^2b = 1.087$, *i.e.* only 8.7% above the relationship $a^2b = 1$ derived by Johansson and George (2006). According to Johansson and George (2006), the virtual origin $x_0$ is supposed to be the same in (10) and (11). This is, however, not satisfied in equations (13) and (14). We notice that the value $x_{02} = 5.135$ in equation (14) is almost exactly twice of $x_{01} = 2.512$ in equation (13). The reason for this is not clear, but we hypothesize that because the main coherent vortex in the present wake originally originates from one vortex sheet separated from one side of the spheroid, that sheet carries only half of the vorticity separated from the spheroid.

## 6. Concluding Remarks

In this Chapter we have discussed the instabilities in the wake behind a 45º inclined 6:1 prolate spheroid, based on direct numerical simulation (DNS) results. We investigated the wake over a wide range of Reynolds numbers ($Re_D$) varying from 10 to 3000.

At low, but not too low, Reynolds numbers ($50 \leq Re_D \leq 800$), the main features of the wake are the vortex sheets in the near-wake that separate from both sides of the spheroid, a counter-rotating vortex pair in the intermediate wake, and the far wake evolved from the vortex sheets. The wake in this Reynolds number range is steady and laminar, no vortex shedding or vortex tube flapping can be observed. Moreover, the vortical structures on both sides of the wake are perfectly symmetric about the symmetry plane, and no side force can be detected on the spheroid.

We identified a primary instability of the wake at $Re_D \approx 1000$. The main features of the wake at $Re_D = 1000$ resemble those at lower Reynolds numbers, although the wake is no longer perfectly symmetric. The near wake, *i.e.* the vortex sheets, is still symmetric, whilst the intermediate wake is clearly asymmetric. A minute unsteadiness, whose amplitude is smaller than 1% of $U_0$ but yet physical, was detected in the intermediate and far wake (further than $14D$ downstream), and believed to be a manifestation of the primary instability. Minute disturbancies induce a loss of wake symmetry, the delicate balance between the two vortex tubes are broken. The weaker vortex tube is heavily deformed by the strain field generated by the stronger vortex tube, which is more concentrated in shape, and both vortex tubes are deflected to the side where the stronger vortex tube resides.

A careful comparison of the wake dynamics at three different Reynolds numbers, $Re_D = 800$, 1000, and 1200, was presented. The comparison showed a clear scenario of the wake transition and loss of symmetry within a narrow Reynolds number band. Though the wake at $Re_D = 1200$ is still only partially asymmetric, the inception of the asymmetry occurs at a more upstream location as compared to the $Re_D = 1000$ wake. The wake at $Re_D = 1200$ is also deflected to one side, as in the $Re_D = 1000$ case, and we found no indication that the deflection would flip to the other side. The side to which the wake first tilts is believed to be arbitrary and a result of a pitchfork bifurcation. In a streamwise slice at $Re_D = 1200$, we observed that the deformed weaker vortex tube periodically rotates around the concentrated stronger vortex tube, which is different from the $Re_D = 1000$ wake in which the strain field generated by the stronger vortex remained steady despite the broken symmetry. These observations again prove that the minute unsteadiness in the $Re_D = 1000$ wake is the initial bifurcation, and the wake is just about to transit at $Re_D = 1000$.

We continued the investigations at the higher $Re_D = 3000$ at which the wake is almost turbulent, but we could still observe the vortex pair in the time-averaged flow field. Both instantaneous and time-averaged flow fields are asymmetric, indicating that the wake, although strongly unsteady and almost fully turbulent, is persistently deflected to one side, similarly as the wakes at $Re_D = 1000$ and 1200. However, the $Re_D = 3000$ wake is significantly different from lower Reynolds number wakes inasmuch as also the near-wake, *i.e.* the separated vortex sheets, is asymmetric. At all lower Reynolds numbers, no matter how asymmetric the intermediate wake is, the near-wake is always symmetric, which is firmly supported by the zero side-force on the spheroid (table 2). The side force at $Re_D = 3000$ is, however, surprisingly high, about 75% of the streamwise drag force. This unexpectedly strong side force was attributed to the heavily distorted mean pressure field around the prolate spheroid. We are inclined to associate the mechanism of this asymmetric wake to a global instability. Anyhow, the observed asymmetric wake, as well as the strong side force, may offer new insight in the symmetry assumption which was widely used in earlier flow simulations of wakes behind prolate spheroids or similar bluff-body geometries, *e.g.* submarine hulls.

There exists a main coherent vortical structure in the complex wake at $Re_D = 3000$. This coherent structure is rather persistent regardless of the complex, chaotic, and turbulent-like small-scale structures surrounding it. By means of the mean helicity density inside the vortex tube, we could confirm that the vortex tube is a helical vortex. In order to investigate this complex coherent vortical structure, we introduced a local cylindrical coordinate system with its axis along the three-dimensional vortex centerline. Rather surprisingly we discovered mechanical energy conservation inside this coherent structure. We moreover identified a new helical symmetry alteration form in this concentrated helical vortex. The vortex tube first has a jet-like axial velocity profile over the vortex cross-section together with positive azimuthal vorticity, *i.e.* a right-handed helical vortex. At a certain location in the intermediate wake, the structure of this helical vortex experiences abrupt changes. The axial velocity profile changes from jet-like to wake-like, while the azimuthal vorticity simultaneously turns negative. These abrupt changes can be associated with a vortex breakdown process. However, based on the classical definition of a vortex breakdown, a recirculation bubble must exist during the breakdown, *i.e.* a reversed axial velocity should appear along the vortex axis. In the present helical vortex tube, no such flow reversal could be seen. Next, by examining the helical pitch of the vortex, we realized that the helical pitch experiences an abrupt change from positive to negative where the change in axial velocity profile and azimuthal vorticity also take place. This indicates that the main coherent structure in the $Re_D = 3000$ wake experienced a helical symmetry alteration. Such a helical symmetry alteration scenario has never been observed in any earlier experiment or simulation. We have therefore, for the first time, reported and analyzed this helical alteration scenario.

Finally, we also examined the validity of a self-similarity law in the $Re_D = 3000$ far-wake. Rather unexpectedly we found that the analytical relationships proposed by Johansson and George (2006) are valid even in this complex wake flow. This may support other and more general applications of these simple self-similarity laws in engineering problems.

**ACKNOWLEDGEMENT**

This work has been supported by the Research Council of Norway through a grant of computing time on the national HPC infrastructure (Programme for Supercomputing, projects nn9191k and nn2469k). F.J. acknowledges the funding from the Future Industry's Leading Technology Development program (No. 10042430) of MOTIE/KEIT of Korea. V.L.O. acknowledges the Russian Science Foundation (grant no. 14-29-00093). We finally thank AIP Publishing and Cambridge University Press for permissions to reprint figure nos 15, 17, 20, and 21.